%% file: arxiv.tex
\documentclass[12pt]{article}

\usepackage[margin=1in]{geometry}

\usepackage{tgtermes}
\usepackage{newtxtext}
\usepackage{newtxmath}
\usepackage{bm}

\usepackage{algorithm}
\usepackage{algpseudocode}
\usepackage{tikz}
\usetikzlibrary{positioning}
\usepackage{longtable}
\usepackage{booktabs}
\usepackage{amsmath}
\usepackage{float}
\usepackage{adjustbox}

\usepackage{xcolor}
\definecolor{darkgreen}{rgb}{0,0.5,0}
\definecolor{darkblue}{rgb}{0,0,0.5}
\definecolor{strcolor}{rgb}{0.6, 0.2, 0.6}
\definecolor{commentcolor}{rgb}{0.3125, 0.5, 0.3125}
\definecolor{keycol}{rgb}{0, 0, 1}

\usepackage{hyperref}
\hypersetup{
    colorlinks=true,
    linkcolor=blue,
    citecolor=darkgreen,
    urlcolor=darkblue,
}
\usepackage{bookmark}

\usepackage{natbib}
\bibpunct[, ]{(}{)}{,}{a}{}{,}%
\usepackage{subcaption}
\usepackage{graphicx}

\usepackage{caption}

\long\def\TABLE#1#2#3{%
  \refstepcounter{table}%
  \begin{minipage}{\textwidth}
  \centering
  \textbf{Table~\thetable.}\quad #1\par\medskip
  \adjustbox{max width=\textwidth}{#2}%
  \par\medskip
  {\small #3}%
  \end{minipage}%
}

\newcommand{\up}{\rule{0pt}{2.5ex}}
\newcommand{\down}{\rule[-1.2ex]{0pt}{0pt}}

\newenvironment{APPENDICES}{}{}

\title{Learning to Pay Attention: Unsupervised Modeling of\\
Attentive and Inattentive Respondents in Survey Data}

\author{
  Ilias Triantafyllopoulos\\[0.3em]
  \normalsize Stern School of Business, New York University\\
  \normalsize New York, NY, USA\\
  \normalsize \texttt{ilias.triantafyllopoulos@nyu.edu}
  \and
  Panos Ipeirotis\\[0.3em]
  \normalsize Stern School of Business, New York University\\
  \normalsize New York, NY, USA\\
  \normalsize \texttt{panos@nyu.edu}
}

\date{}

\begin{document}
%% ---------------------------------------------------------------

\maketitle

\begin{abstract}
\input{sections/abstract}
\end{abstract}

\input{sections/introduction}

\input{sections/related_work}

\input{sections/methods}

\input{sections/experiments}

\input{sections/analysis}

\input{sections/implications}

\input{sections/conclusions}

\input{sections/limitations}

\bibliographystyle{plainnat}
\bibliography{references}

\clearpage
\appendix
\input{sections/appendices}

\end{document}

%% file: sections/abstract.tex
The integrity of behavioral and social-science surveys depends on detecting inattentive respondents who provide random or low-effort answers. Traditional safeguards, such as attention checks, are often costly, reactive, and inconsistent. We propose a unified, label-free framework for inattentiveness detection that scores response coherence using complementary unsupervised views: geometric reconstruction (Autoencoders) and probabilistic dependency modeling (Chow-Liu trees). While we introduce a "Percentile Loss" objective to improve Autoencoder robustness against anomalies, our primary contribution is identifying the structural conditions that enable unsupervised quality control. Across nine heterogeneous real-world datasets, we find that detection effectiveness is driven less by model complexity than by survey structure: instruments with coherent, overlapping item batteries exhibit strong covariance patterns that allow even linear models to reliably separate attentive from inattentive respondents. This reveals a critical ``Psychometric-ML Alignment'': the same design principles that maximize measurement reliability (e.g., internal consistency) also maximize algorithmic detectability. The framework provides survey platforms with a scalable, domain-agnostic diagnostic tool that links data quality directly to instrument design, enabling auditing without additional respondent burden.

%% file: sections/introduction.tex
\section{Introduction}\label{sec:Intro}

\begin{figure}[htbp]
    \centering
    
    \includegraphics[width=\linewidth]{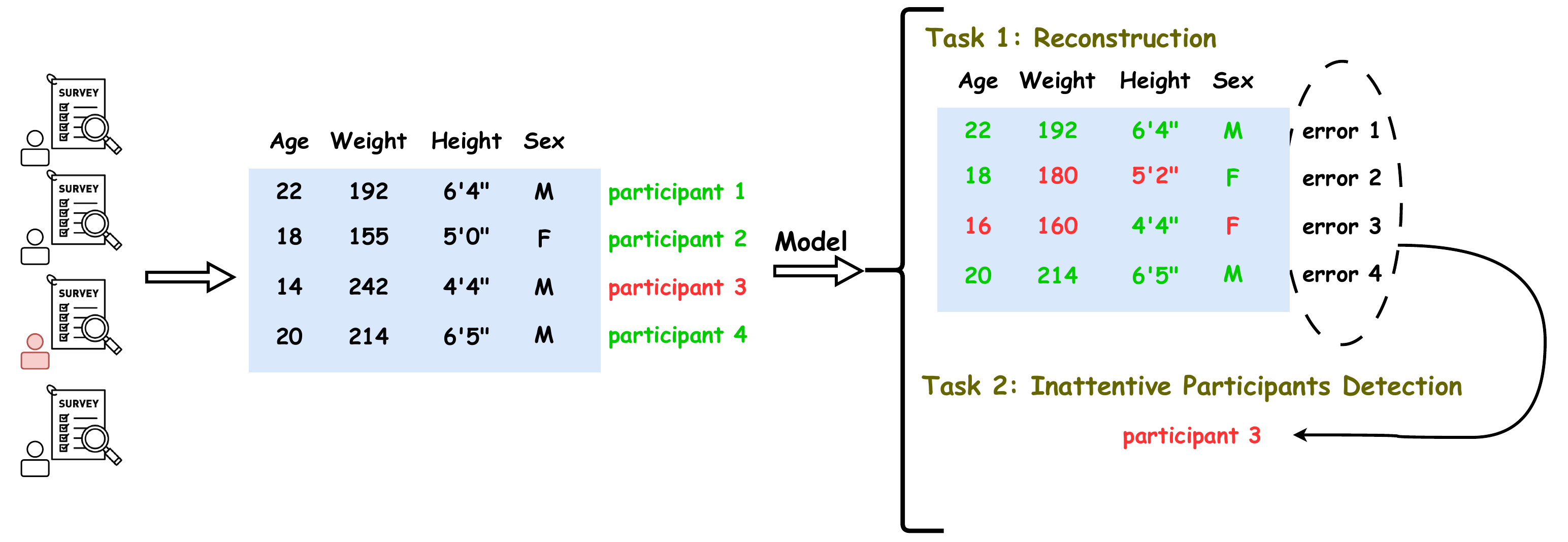}
    \caption{A toy example illustrating the proposed approach.
    Survey responses from multiple participants are represented as structured categorical data across questions (e.g., $\langle \textit{Age}, \textit{Height}, \textit{Sex}, \textit{Weight} \rangle$).
    The model first learns to reconstruct each participant’s responses (Task 1) and, based on reconstruction errors, detects incoherent (hard-to-model) responses.
    We hypothesize that participants with many such inconsistencies are inattentive (Task 2). Green denotes attentive respondents or well-reconstructed answers;
    red indicates inattentive respondents or poorly reconstructed patterns. Together, the two tasks demonstrate how unsupervised reconstruction enables automated detection of low-quality survey data.}
    \label{fig1_archi}
\end{figure}

Behavioral, social, and political scientists rely heavily on surveys to capture human sentiment and behavior.
However, the validity of this research is constantly threatened by ``content nonresponsivity''—responses that fail to consistently reflect the respondent's internal state due to inattention, fatigue, or lack of engagement.
Researchers define this phenomenon as a loss in the consistency of responses to data items~\citep{nichols1989}.
Respondents may provide random answers due to survey fatigue or unwillingness, thereby affecting the validity of research outcomes~\citep{meade2012}.
This issue has been exacerbated by the growing reliance on online crowdsourcing platforms such as Amazon Mechanical Turk (MTurk) and Prolific.
While these platforms offer fast and diverse data collection, workers are often more likely to multitask during studies~\citep{goodman2013}, and inattentive or fraudulent responses are more common, particularly when participants have the opportunity to cheat~\citep{peer2022}.

Current techniques for mitigating these issues remain largely reactive. Researchers rely on embedded attention checks~\citep{berinsky2014}, metrics of individual response variation~\citep{curran2016}, or mechanism-design approaches like the Bayesian Truth Serum~\citep{weaver2013}.
While useful, these methods impose significant ``taxes'' on the research process: they increase respondent cognitive load, extend survey completion time, and can induce measurement reactivity.
% Furthermore, they are often binary and rigid, failing to capture the spectrum of engagement that varies across a long instrument.

Unsupervised machine learning offers a promising alternative: detecting inattentiveness by modeling the ``coherence'' of response patterns directly, without requiring labeled training data or explicit ``trap'' questions.
Crucially, this approach addresses a fundamental limitation of supervised classification in this domain: the impossibility of objective ground truth.
When collecting respondents' data, we cannot definitively ensure who answered randomly or if they gave fabricated information, as we do not have a ground truth to compare against.
Creating datasets that include this information would require substantial cost and time, and even if achievable, they would likely include biases and inconsistencies, as definitions of inattentiveness vary across studies.
Consequently, unsupervised learning techniques are not just efficient; they offer a necessary, label-free alternative to ground-truth dependence.

In this study, we move beyond simple algorithm benchmarking to develop a \emph{structural theory of unsupervised data quality}.
We unify and evaluate three domain-agnostic method families: (i) non-linear autoencoders (AE), (ii) linear (zero-layer) autoencoders, and (iii) generative Bayesian networks with Chow–Liu tree structure~\citep{chow1968}.
To our knowledge, no prior work benchmarks these families side-by-side for inattentiveness detection in structured survey data.
Furthermore, we empirically characterize a key phenomenon: as an AE’s reconstruction capacity increases to fit all inputs, anomaly sensitivity can diminish because even incoherent patterns begin to be reconstructed well.
To address this \emph{reconstruction-detection trade-off} and the challenge of ``contaminant learning'' (where models learn to reconstruct noise), we introduce a \emph{Percentile Loss} (PL) objective, adapting a robust training technique to the survey domain.

This label-free approach is critical for scalable deployment. First, it avoids measurement reactivity and reduces the risk of automation.
Second, it remains compatible with legacy datasets that lack embedded checks.
Third, it lowers dependence on context-specific rules, enabling pre-deployment screening and post-collection auditing across heterogeneous samples~\citep{hong2020}.

Our contributions are the following:
\begin{enumerate}
    \item \textbf{Comprehensive Benchmarking on ``Uncleaned'' Data:} Validating unsupervised detection requires raw datasets where inattentive responses have \emph{not} yet been filtered out—a rarity in public repositories where data is typically pre-cleaned. We curate nine heterogeneous real-world datasets spanning diverse populations (adolescents, MTurk workers, representative samples) and topics. This allows us to establish the first rigorous, domain-agnostic benchmark for inattentiveness detection in the absence of perfect ground truth.

    \item \textbf{The Psychometric-ML Alignment:} Using this diverse testbed, we demonstrate that the success of unsupervised detection is governed primarily by \emph{survey structure}, not dataset size or model complexity. We find the performance of various unsupervised models for detecting inattention correlates with well-designed surveys that are built on covariance structures (e.g., coherent, overlapping item batteries)~\citep{cronbach1951,clark1995,clark2019}. This creates a powerful design principle: instruments with high  ``coherence'' are inherently easier to quality-control algorithmically.

    \item \textbf{Robust Percentile Loss (PL):} We introduce \emph{Percentile Loss} for autoencoders in this domain. Instead of minimizing average reconstruction error, PL minimizes the average of the lowest-error $p$-percentile. This resolves the reconstruction-detection trade-off by focusing on the majority structure while de-emphasizing incoherent ``noisy'' cases, preventing the autoencoder from overfitting to anomalies.
    
    \item \textbf{Probabilistic Baseline:} We adapt a Chow–Liu tree to categorical survey data to use its per-row likelihood as a typicality score. While widely used for modeling, we document its competitiveness as an interpretable, unsupervised inattentiveness detector that complements reconstruction-based views.
    
    \item \textbf{Actionable Framework:} We translate these findings into concrete recommendations for designing surveys and setting an adaptable pipeline that can be embedded into organizational survey platforms.
\end{enumerate}

Figure~\ref{fig1_archi} illustrates our two-stage workflow: (1)~unsupervised modeling that reconstructs or probabilistically explains typical patterns, followed by (2)~respondent ranking using reconstruction error or (negative) likelihood for inattentiveness detection.
Our methods are fully label‑free at training time and require no additional items or attention checks, making them practical for existing datasets and for pre‑deployment quality monitoring.

%% file: sections/related_work.tex
\section{Related Work}\label{sec:RW}

\subsection{Survey Inattentiveness Detection}
Traditional approaches to inattentiveness rely on embedded design features such as attention checks, response time thresholds, or pattern-based indicators like straightlining~\citep{kim2019straightlining}. 
The Bayesian Truth Serum~\citep{weaver2013} represents another strand, incentivizing truthful responses through mechanism design.
More advanced statistical and machine learning approaches include EM-based truth discovery~\citep{dawid1979maximum}, unsupervised credibility modeling~\citep{yin2007truth}, and supervised classification using behavioral features~\citep{schroeders2022detecting, ozaki2024detecting}. 
However, supervised methods depend heavily on scarce and survey-specific labels, often derived from attention checks that capture only partial inattentiveness. This limits their scalability and generalization, motivating unsupervised, data-driven alternatives that infer respondent coherence directly from response structure.

Psychometric approaches grounded in Item Response Theory (IRT) provide person-level screening via person-fit statistics, which flag response patterns that are improbable under a fitted model. A prominent example is the standardized log-likelihood index $l_z$, which evaluates how surprising a respondent's full item-score vector is given estimated item parameters and the respondent's estimated latent trait~\citep{drasgow1985appropriateness}. In this context, large negative values indicate careless responding. Subsequent research introduced corrected variants, such as $l_z^*$, to account for the statistical bias that occurs when using estimated ability levels rather than true latent traits~\citep{snijders2001asymptotic}.
When full IRT calibration is computationally or practically impractical, simpler "group-based" proxies approximate the same intuition by checking whether a respondent's answers align with empirical item easiness (i.e., the average response rate or mean score for each item across the sample). For instance, the person-total correlation $r_{pt}$ computes the correlation between an individual's response vector and the vector of item means ordered by difficulty~\citep{donlon1968index}. Near-zero or negative values for $r_{pt}$ suggest that the respondent is answering randomly or in a manner that is incoherent relative to the rest of the population.
% These methods closely relate to our framing of inattentiveness as low \emph{typicality}: like $l_z$ they operationalize respondent quality through improbability of the full response pattern, while our autoencoder reconstruction losses and Chow--Liu likelihood scores estimate typicality directly from learned cross-item regularities without requiring a unidimensional IRT specification.

\subsection{Autoencoders for Anomaly Detection \& Survey Inattentiveness}
Autoencoders were originally introduced for unsupervised anomaly detection because they can reconstruct typical inputs while producing higher errors for unusual ones~\citep{hawkins2002outlier}. This principle has been extended with variants such as Denoising Autoencoders~\citep{vincent2010stacked}, Variational Autoencoders~\citep{kingma2013auto}, and Robust Deep Autoencoders~\citep{zhou2017anomaly}, each aiming to improve robustness to noise or explicitly model outliers. These innovations established autoencoders as a flexible foundation for anomaly detection in diverse domains~\citep{goodge2021robustness, rubio2020classifying}.

Only recently have autoencoders been applied specifically to inattentive responding. \cite{alfons2024open} introduced autoencoders for classifying inattentive responses, but their evaluation was limited to synthetic data. \cite{welz2023respondents} proposed CODERS, which combines autoencoder reconstruction errors with changepoint detection to identify when a respondent begins answering carelessly within a survey, demonstrating proof-of-concept in one real and several synthetic datasets. Our work differs in three key respects. First, we adapt autoencoders for categorical survey data, introducing variable-level loss weighting and Percentile Loss to address the reconstruction–detection trade-off. Second, instead of focusing on within-survey onset, we target the identification of inattentive respondents as a whole. Third, we validate our approach on nine diverse, real-world survey datasets that include both attentive and inattentive participants, establishing large-scale empirical benchmarks. In doing so, we position autoencoders as a scalable and domain-agnostic tool for inattentiveness detection in survey practice.

\subsection{Probabilistic Methods}
In recent years, probabilistic models have been leveraged for detecting anomalies or inattentive respondents in survey data. Bayesian networks can approximate the joint distribution of item responses, flagging outliers~\citep{lindskou2021detecting}. Latent class analysis and mixture modeling approaches similarly identify outliers by positing an “outliers” latent class separate from normal. For example, mixture IRT models distinguish a class of insufficient-effort respondents (characterized by near-random or content-independent answers) from attentive respondents, improving parameter estimation and person-fit~\citep{ulitzsch2024screen}.  The probabilistic frameworks have also demonstrated practical utility in even detecting automated “bot” respondents via Bayesian latent-class joint models~\citep{roman2022automated}.

\subsection{Survey Design and Detection Reliability} The use of overlapping or coherent item batteries have a substantial impact on the effectiveness of inattentiveness detection. Psychometric theory holds that redundant, correlated items provide built-in consistency checks that can reveal careless responding. 
When multiple items measure the same construct (e.g. an item battery of related questions), attentive respondents should answer them in a consistent manner, yielding a strong shared variance and high internal reliability. 
By contrast, a careless or random responder will often disrupt this covariance pattern by answering related items inconsistently. 
Empirically, the presence of an expected correlation structure makes it easier to spot outliers. For example, psychometric synonym or antonym pairs (nearly overlapping items) have been used to check consistency: a respondent who gives divergent answers to two questions that should elicit similar answers is likely not paying attention~\citep{meade2012, goldammer2020careless}. If each item were unique (measuring entirely distinct constructs), such checks would be far less powerful because there is no expected pattern to violate. 
Thus, researchers emphasize careful survey design, constructing coherent item batteries with adequate breadth and redundancy, as a means to improve the reliability of the data~\citep{clark2019}, and (we are arguing) the detection of inattentive responders.

\section{Data}\label{sec:Data}
We describe the datasets that we used for our study and give the main points of each dataset, along with how the data were collected and what attention checks were included. In Table~\ref{tab:data_stats}, we summarize the statistics of all datasets, including the number of samples, the number of variables, the number of features, and the average number of features per variable.
We used \url{https://datasetsearch.research.google.com/} to find publicly available survey datasets that include attention checks and use mainly structured/categorical responses (as opposed to textual or other forms of unstructured data). We identified nine datasets that differ substantially in topic, respondent population, and quality-control mechanisms to ensure that the evaluation reflects the method’s robustness across diverse contexts. This heterogeneity spans (1) respondent types (such as adolescents~\citep{robinson2014}, MTurk workers~\citep{moss2023}, and nationally representative adult samples~\citep{mastroianni2022}; (2) survey topics (ranging from political attitudes to misinformation susceptibility); and (3) attention check designs (ranging from none~\citep{robinson2014} to multiple embedded checks~\citep{pennycook2020}.
Beyond diversity, we applied two inclusion criteria: (a)~datasets had to contain attention checks (or clear knowledge checks on when someone answers inattentively for the case of~\cite{robinson2014}), and (b)~they had to retain participants' responses who failed these checks. These criteria were highly exclusive because most published works release only ``cleaned'' datasets with inattentive respondents removed, making large-scale evaluation of inattentiveness detection challenging.

\begin{table}
\TABLE
{Summary of Datasets \label{tab:data_stats}}
{\begin{tabular}{@{}l@{\quad}cccc@{}}
        \textbf{Dataset} & \textbf{Samples} & \textbf{Variables} & \textbf{Features} & \textbf{AFV} \\

\hline\up 
\cite{robinson2014} & 14,765 & 98 & 619 & 6.32 \\
        \cite{pennycook2020} & 853 & 188 & 708 & 3.75 \\
        % \quad Condition 1 & 212 & 98 & 404 & 4.12 \\
        % \quad Condition 2 & 206 & 98 & 358 & 3.65 \\
        % \quad Condition 3 & 220 & 98 & 376 & 3.84 \\
        % \quad Condition 4 & 215 & 98 & 400 & 4.08 \\
        \cite{alvarez2019} & 2,725 & 39 & 196 & 5.03 \\
        \cite{uhalt2020} & 308 & 60 & 337 & 5.62 \\
        \cite{ogrady2019} & 355 & 72 & 322 & 4.47 \\
        \cite{buchanan2018} & 1,038 & 23 & 159 & 6.91 \\
        \cite{moss2023} & 2,277 & 51 & 332 & 6.51 \\
        \cite{mastroianni2022} & 1,036 & 51 & 322 & 6.31 \\
        \cite{ivanov2021} & 860 & 67 & 310 & 4.63 \\ \hline
\end{tabular}}{Every sample consists of a number of variables. Variables can be either questions or demographic records. Every variable consists of features in the way it was explained in Section~\ref{data_preparation}. The Average number of Features per Variable (AFV) is also given in the last column.}
\label{datasets_stats}
\end{table}

\textbf{\cite{robinson2014}}, \textbf{Mischievous Respondents:} This dataset comes from the 2012 Dane County Youth Assessment (DCYA), an anonymous web-based survey of 14,765 U.S. high school students. The study investigated “mischievous responders”, adolescents who intentionally gave extreme or implausible answers (e.g., exaggerated reports of health behaviors) that distort between-group disparity estimates across categories such as sexual orientation, gender identity, and disability. Outcomes examined included suicidal ideation, school belongingness, and substance use. Unlike other datasets in our study, this survey contained no embedded attention checks; inattentive cases were instead identified through domain knowledge and responses to unrelated questions. This makes the dataset distinctive, as it provides a large-scale setting where inattentiveness is inferred from patterns of extreme or inconsistent responding rather than explicit screening items.

\textbf{\cite{pennycook2020}}, \textbf{COVID-19 Misinformation:} This study examined how attention and cognitive reflection affect the spread of COVID-19 misinformation. A total of 853 U.S. respondents evaluated 30 headlines (15 true, 15 false) presented in a social media format, with ground truth verified by fact-checking sources. The central aim was to test whether misinformation sharing occurs because people fail to consider accuracy rather than because they truly believe false claims. To measure this, the study introduced four experimental conditions that varied both the framing of the question (accuracy vs. sharing intention) and the ordering of response options. Data quality was assessed with multiple attention checks: two multiple-choice instruction items, one embedded Likert-scale item requiring selection of a specific value (“3”), and an additional self-report item asking whether participants had responded randomly. This combination of structured conditions and varied attention checks makes the dataset particularly useful for evaluating inattentiveness detection under different operational definitions.

\textbf{\cite{alvarez2019}}, \textbf{Inattentive:} This survey collected 2,725 responses from California adults through Qualtrics’ e-Rewards panel to study how inattentiveness affects political attitude measures. The study examined whether inattentive respondents introduce noise, satisficing, or misreporting, and whether they differ demographically from attentive participants. Data quality was monitored with three trap questions: two multiple-choice items requiring a specific answer and one open-text item requiring the word “government.” Importantly, the full dataset includes both attentive and inattentive respondents, making it a strong benchmark for evaluating detection methods in a domain with coherent, multi-item political constructs.

\textbf{\cite{uhalt2020}}, \textbf{Attention Checks and Response Quality:} This Qualtrics survey included 308 participants and focused on self-assessment of personality traits using Likert-scale items. To evaluate response quality, the instrument embedded multiple explicit attention checks that instructed participants to select specific options (e.g., “I see myself selecting ‘Agree Strongly’ if I’m paying attention to the survey”). These checks were distributed throughout the questionnaire to monitor engagement. The dataset provides a useful test case because it combines a structured personality scale, where attentive responses should show internal consistency, with clear ground-truth labels from embedded checks.

\textbf{\cite{ogrady2019}} \textbf{Moral Foundations:} This study collected 355 valid responses from U.S. undergraduate business students to examine how moral foundations predict prosocial behavior. Participants completed the Moral Foundations Questionnaire (MFQ), which measures individualizing foundations (care, fairness) and binding foundations (loyalty, authority, purity). Data quality was monitored with a multiple-choice instructional manipulation check (IMC) embedded in the survey to ensure that respondents read instructions carefully. The dataset offers a structured attitudinal battery with a single attention check for identifying inattentiveness.

\textbf{\cite{buchanan2018}}, \textbf{Low-Quality Data:} This online Qualtrics survey collected 1,038 valid responses and examined methods for detecting low-quality data in psychological research. The study explored inattentive, low-effort, and automated responses (e.g., survey bots) using behavioral and statistical indicators such as response times, click counts, and distributional anomalies. Data quality was also monitored with an embedded attention check, instructing participants to “Please mark strongly agree for this question.” The dataset is distinctive in combining traditional survey responses with metadata-based quality indicators, making it useful for testing detection methods that rely solely on response patterns.

\textbf{\cite{moss2023}}, \textbf{Ethical Data:} Study 1 surveyed 2,277 active U.S.-based MTurk workers to examine financial dependence on MTurk, time investment, and perceptions of fairness. The dataset is heterogeneous, covering multiple facets of workers’ experiences. Data quality was monitored with two screens: one required selecting the correct summary of a prior item, while the other directly asked participants to indicate agreement with the statement “I am not reading the questions in this survey.” This combination provides both indirect and self-report measures of inattentiveness.

\textbf{\cite{mastroianni2022}}, \textbf{Attitude Change:} Study 1 collected 1,036 responses from a nationally representative U.S. adult sample via Prolific to examine perceptions of long-term societal attitude change. Participants compared their estimates of historical public opinion trends to actual polling data, with particular interest in whether people systematically overestimate liberalization. Data quality was ensured with an embedded attention check requiring respondents to type the number “1” in both blanks of a survey item. This dataset provides a high-quality, representative benchmark with a single IMC for inattentiveness.

\textbf{\cite{ivanov2021}} \textbf{Racial Resentment:} This MTurk survey collected 860 responses in April 2020 to study attitudes toward decarceration during COVID-19. The study examined how information about prison health risks, racial resentment, and empathy influenced support for release, and whether support varied by crime type, age, or health status of incarcerated individuals. Data quality was monitored with two embedded attention checks. The dataset offers a heterogeneous attitudinal instrument with multiple checks, useful for assessing inattentiveness in politically sensitive contexts.

%% file: sections/methods.tex
\section{Methods}\label{sec:Methodology}

Our study investigates unsupervised domain-agnostic ML approaches for detecting inattentive survey respondents. We compare three families of methods: (i) non-linear autoencoders enhanced with a targeted percentile loss, (ii) a basic linear autoencoder with zero hidden layers, and (iii) a probabilistic Bayesian Network. All methods are applied to structured categorical variables and categorized numerical variables.

\subsection{Data Preprocessing}
For our experiments, we focus exclusively on categorical variables. However, we also take into consideration the numeric variables: if they have fewer than 20 distinct values, we treat them as categorical. Otherwise, we discretize them into predefined categories based on their standardized values. First, we apply standard normalization to each numeric variable $z = (x - \mu) / \sigma$ where x represents the raw value, $\mu$ is the mean, and $\sigma$ is the standard deviation of the variable. Afterwards, we categorize the values into six discrete bins, as defined in Table~\ref{tab:dicretization}.

\begin{table}
\TABLE
{Numeric Variables Categorization \label{tab:dicretization}}
{\begin{tabular}{@{}l@{\quad}c@{}}
        \textbf{Category} & \textbf{Standardized Value Range} \\
\hline\up 
        Bottom-extreme  & $z < -1.4$  \\
        Low            & $-1.4 \leq z < -0.7$ \\
        Normal         & $-0.7 \leq z \leq 0.7$ \\
        High           & $0.7 < z \leq 1.4$ \\
        Top-extreme    & $z > 1.4$  \\
        Missing Data   & N/A  
        \down\\ \hline
\end{tabular}}{Categorization of numeric variables based on standardized values.}
\end{table}

Admittedly, many questions were open-ended and required text as a response. We focus exclusively on structured categorical and categorized numerical variables, thereby excluding open-ended responses from our analysis. We leave the analysis of surveys with open-ended responses as a fruitful direction for future research.

\subsection{Data Preparation}
\label{data_preparation}
The datasets we use in this study are surveys, where each participant provides answers to a series of usually multiple-choice questions. Since the majority of survey items follow a closed-form structure, the data is inherently categorical. All our methods function with numeric data, and thus, we transform the categorical variables into a numerical representation proper for the network.  Each survey question (variable) is encoded using one-hot encoding, where each possible response choice is represented as a separate binary feature. Specifically, for a question with k possible answer choices, a vector of length k is created (or k+1 if there exist users who passed the question), where only the selected response is marked as 1 (the remaining elements are set to 0). This transformation results in a feature space where the number of input dimensions is significantly larger than the number of original survey items, as each categorical variable expands into multiple binary features. Each model treats each response as a set of features, learning the underlying distribution of the entire response set. 

\subsection{Non-Linear Autoencoders}

\begin{figure}[htbp]
    \centering
    \includegraphics[width=0.6\linewidth]{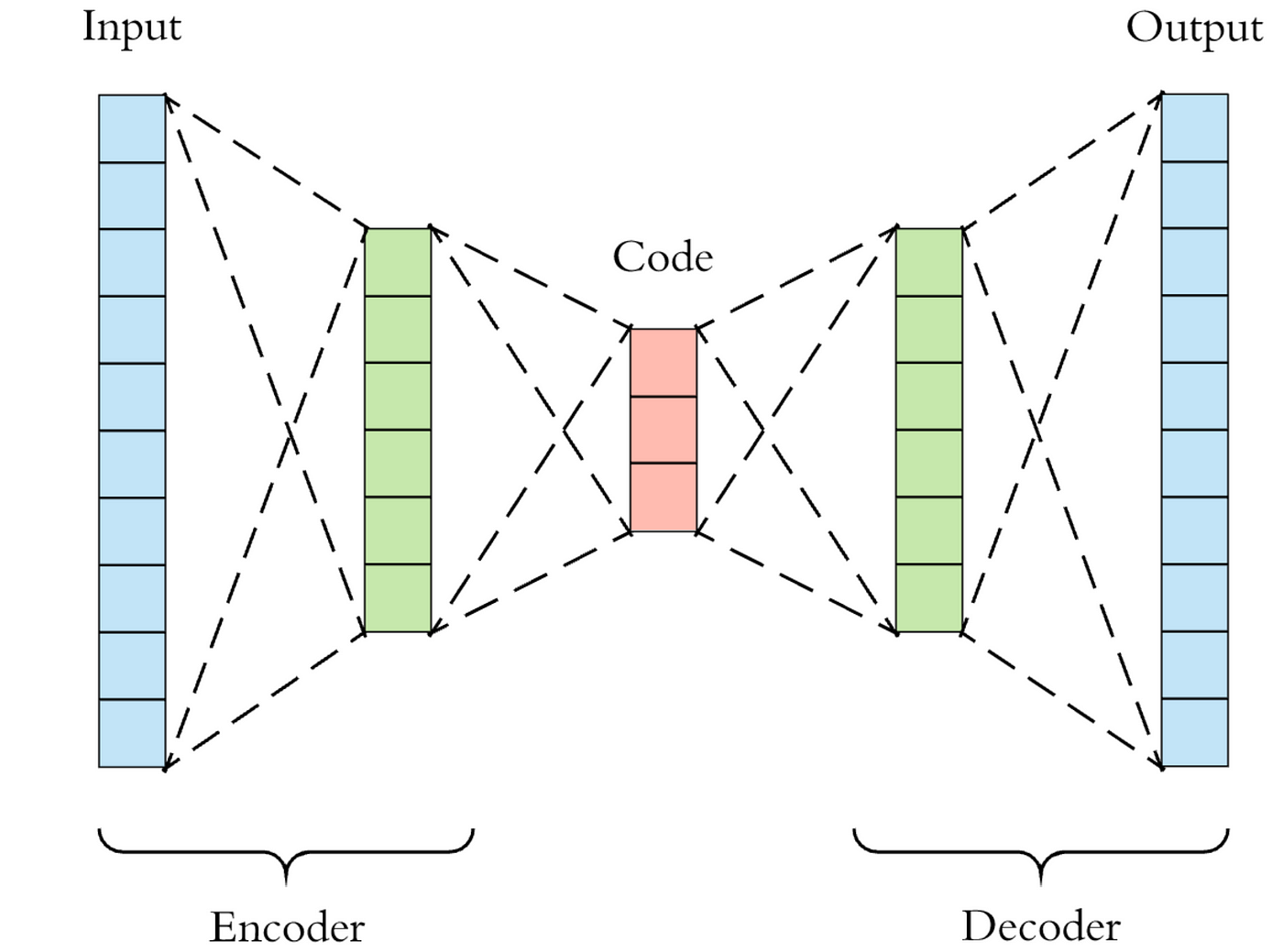}
    \caption{Architecture of a simple Autoencoder. The autoencoder consists of two parts: (a) the Encoder, which encodes the information into latent variables, and (b) the Decoder, which decodes the information to the initial input. \href{https://www.digitalocean.com/community/tutorials/autoencoder-image-compression-keras}{Source}}
    \label{autoencoders_archh}
\end{figure}

Autoencoders are a class of neural networks designed to learn compact representations of data in an unsupervised manner by encoding inputs into a lower-dimensional latent space and reconstructing them from this compressed representation. An autoencoder consists of two parts: an encoder, which learns an efficient representation of the data in a lower dimension, and a decoder, which learns how to create the initial data from this low-dimensional representation (Figure~\ref{autoencoders_archh}).

\paragraph{Encoder:} The encoder function, denoted as $f_{\theta}$, maps an input response vector $\mathbf{x} \in \mathbb{R}^{d}$ (where $d$ is the number of features) to a lower-dimensional latent representation $\mathbf{z} \in \mathbb{R}^{m}$ (with $m \ll d$):
\begin{equation*}
    \mathbf{z} = f_{\theta}(\mathbf{x}) = \sigma(W_e \mathbf{x} + \mathbf{b}_e)
\end{equation*}
where $W_e \in \mathbb{R}^{m \times d}$ is the weight matrix, $\mathbf{b}_e \in \mathbb{R}^{m}$ is the bias term, and $\sigma(\cdot)$ is a non-linear activation function such as ReLU or sigmoid.

\paragraph{Decoder:} The decoder function, denoted as $g_{\phi}$, reconstructs the original response vector from the latent representation:
\begin{equation*}
    \hat{\mathbf{x}} = g_{\phi}(\mathbf{z}) = \sigma(W_d \mathbf{z} + \mathbf{b}_d)
\end{equation*}
where $W_d \in \mathbb{R}^{d \times m}$ is the decoder weight matrix, $\mathbf{b}_d \in \mathbb{R}^{d}$ is the bias term, and $\hat{\mathbf{x}} \in \mathbb{R}^{d}$ is the reconstructed response vector.

The number of latent variables, $m$, as well as the depth of the encoder and decoder, the number of units per layer, and the choice of activation functions are all hyperparameters that can be tuned for optimal performance. To prevent overfitting and enhance generalization, L2 regularization (also known as weight decay) is applied to the weights of the network \cite{krogh1991}. The regularized objective function is given by:
\begin{equation*}
\mathcal{L}_{\text{reg}} = \mathcal{L}(\mathbf{x}, \hat{\mathbf{x}}) + \lambda \sum ||W||_2^2
\end{equation*}
where $\lambda$ is a regularization hyperparameter that controls the penalty on large weight values.

Additionally, dropout is employed in hidden layers to randomly deactivate neurons during training \cite{srivastava2014}. During training, each unit is dropped with a probability $p_{\text{drop}}$:
\begin{equation*}
\mathbf{h}^{(l)}{\text{drop}} = \mathbf{r} \odot \mathbf{h}^{(l)}, \quad r_i \sim \text{Bernoulli}(1 - p_{\text{drop}})
\end{equation*}
where $\mathbf{h}^{(l)}$ is the activation vector in layer $l$, and $\odot$ denotes element-wise multiplication with the dropout mask $\mathbf{r}$. $p_{\text{drop}}$ is also a hyperparameter to be tuned.

\paragraph{Loss Function.}

For categorical survey responses, the Binary Cross-Entropy (BCE) loss is used for training the autoencoders:
\begin{equation*}
    \text{BCE}(\mathbf{x}, \hat{\mathbf{x}}) 
    = - \sum_{i=1}^{d} \Big[ x_i \log \hat{x}_i + (1 - x_i) \log (1 - \hat{x}_i) \Big].
\end{equation*}

As explained earlier, our datasets consist of variables (questions), each of which expands into multiple binary features after one-hot encoding. To respect this nested structure, we modify the BCE loss to compute the error per variable and normalize by the logarithm of the number of features, so that questions with many response options do not disproportionately dominate the total loss. Formally:
\begin{equation*}
    \mathcal{L}_{\text{base}} 
    = \frac{1}{|\mathcal{V}|} 
    \sum_{v \in \mathcal{V}} 
    \frac{1}{\log |F_v|} 
    \sum_{f \in F_v} 
    \text{BCE}(x_f, \hat{x}_f),
\end{equation*}
where $\mathcal{V}$ is the set of survey variables (questions) and $F_v$ is the set of features (possible responses) for variable $v$.
\noindent

While this base loss encourages faithful reconstruction of all input patterns, autoencoders are by design optimized to minimize the overall reconstruction error. This can be suboptimal for our goal of inattentive-user detection: we are not primarily interested in reconstructing every response equally well, but rather in modeling the consistent and majority patterns of the data, while letting incoherent patterns stand out with higher error. To this end, we introduce the \textit{Percentile Loss (PL)} (\cite{merrill2020}) as an extension of the base formulation. PL was originally proposed in the computer vision domain, where the goal was to model the common background rather than rare anomalies. The rationale behind PL is that it is statistically improbable for the 95th-percentile loss in a batch to correspond to an anomaly, assuming anomalies are rare. Accordingly, PL focuses learning on the lowest-error subset of a batch, thereby discouraging the model from overfitting to anomalous inputs. 

\noindent
Let $\ell_j$ denote the total reconstruction loss of sample $j$ under $\mathcal{L}_{\text{base}}$. For a mini-batch of $N$ samples, we sort these losses in ascending order:
\[
    \ell_{(1)} \leq \ell_{(2)} \leq \cdots \leq \ell_{(N)}.
\]
Given a chosen percentile $p \in (0,100]$, we retain only the lowest-error subset of size 
\(
    k = \Big\lfloor \tfrac{p}{100} \cdot N \Big\rfloor,
\) 
ignoring the highest-error samples. The Percentile Loss is then defined as
\begin{equation*}
    \mathcal{L}_{\text{PL}} 
    = \frac{1}{k} \sum_{j=1}^{k} \ell_{(j)}.
\end{equation*}

\noindent
In practice, setting $p < 100$ shifts the learning focus towards accurately reconstructing the bulk of the data distribution, while deliberately de-emphasizing atypical responses. This has the desired effect in our context: attentive respondents, who produce consistent patterns, shape the model’s representation, whereas inattentive respondents, who answer carelessly or randomly, yield disproportionately higher reconstruction errors and can thus be more easily flagged as anomalies. 

% \noindent
% This enriched formulation represents a central contribution of our work: by adapting the objective function from pure reconstruction to selective reconstruction, we explicitly align the training of autoencoders with the ultimate goal of inattentive-user detection.

\paragraph{Randomness Detection Protocol:}

As already explained, autoencoders are suitable models for capturing the internal correlations between features in the dataset and constructing a structured latent representation that preserves these dependencies. Given their ability to learn compact and meaningful representations of valid response patterns, we hypothesize that an autoencoder can effectively differentiate between attentive and inattentive respondents based on reconstruction loss.

A critical methodological distinction in anomaly detection is between \textit{inductive} (detecting new anomalies in future data) and \textit{transductive} (detecting anomalies within the current dataset) settings. Survey auditing is typically transductive: a researcher collects a batch of $N$ responses and aims to clean \textit{that specific batch}. Therefore, our primary evaluation trains the autoencoder on the full dataset and computes reconstruction errors for the same samples. While this ``train-on-test'' approach would be invalid for supervised classification, it is standard in unsupervised outlier detection because the autoencoder is regularized (via the bottleneck $m \ll d$ and Percentile Loss) to capture the \textit{majority manifold}. Anomalies, by definition, fail to fit this manifold even when present in the training set~\citep{zhou2017anomaly, eduardo2020}.

% \emph{TODO AND VERIFY:} To rigorously verify that our models are not simply ``memorizing'' the noise, we also implemented a \textbf{5-fold cross-validation} procedure as a robustness check. In this setup, the model is trained on 80\% of the data (using Percentile Loss to ignore the highest-error samples during training) and reconstruction error is computed on the held-out 20\%. We found that the ranking of inattentive respondents remained stable (rank correlation $\rho > Y.YY $) between the full-batch and cross-validated scores. Thus, for simplicity and maximum data utilization, we report results from the full-batch training in Table~\ref{tab:XXX}.

Therefore, we rank the reconstruction errors in descending order, identifying the respondents with the highest losses as potential inattentive participants. Our hypothesis is based on extensive prior research in anomaly and outlier detection, where autoencoders and similar deep learning architectures have been successfully employed to identify data points that deviate from expected patterns; in Computer Vision~\citep{an2015, zhou2017anomaly}, time series~\citep{xu2018unsupervised}, and tabular data~\citep{eduardo2020}.

Our approach builds on the assumption that inattentive respondents lack consistent response patterns, producing randomness that the autoencoder cannot reconstruct. By contrast, minority groups with coherent but distinctive perspectives remain reconstructible because their internal consistency anchors them within the learned manifold. This distinction mirrors earlier dimensionality reduction techniques such as Principal Component Analysis (PCA)~\citep{abdi2010}, where structured minority patterns are captured by weaker components, whereas purely random responses cannot be represented. Thus, the autoencoder primarily flags incoherence rather than legitimate minority viewpoints, clarifying an important boundary condition of our method.

\paragraph{Setup:}
\label{setup}

For each dataset (survey), we separately perform a hyperparameter tuning phase before training the autoencoder. We employ  Bayesian Optimization to efficiently explore the hyperparameter space. This optimization is conducted using the KerasTuner package within the Keras framework. Table~\ref{tab:hyperparameter} provides an overview of the hyperparameters explored during tuning, along with their respective search ranges. The optimization process evaluates various configurations over 30 trials, using a validation split of 20\% and training each candidate model for up to 300 epochs, with an Early Stopping of 10 epochs. The best-performing configuration is selected based on validation loss minimization.

\begin{table}
\TABLE
{Hyperparameter search space \label{tab:hyperparameter}}
{\begin{tabular}{@{}l@{\quad}c@{}}
        \textbf{Hyperparameter} & \textbf{Values Explored} \\
\hline\up 
        Learning Rate & \{0.0001, 0.001, 0.01\} \\
        Encoder Layers & \{1, 2, 3\} \\
        Encoder Units & \{64, 96, 128, 160, 192, 224, 256\} \\
        Encoder Activation & \{ReLU, SeLU, GeLU, Swish\} \\
        Encoder Regularization & \{0.0, 0.001, 0.01\} \\
        Encoder Dropout & \{0.0, 0.1, 0.2, 0.3, 0.4, 0.5\} \\
        Encoder Batch Normalization & \{True\} \\
        Latent Space Dimensionality & \{2, 3, ..., 50\} \\
        Latent Activation & \{ReLU, SeLU, GeLU, Swish\} \\
        Decoder Layers & \{1, 2, 3\} \\
        Decoder Units & \{64, 96, 128, 160, 192, 224, 256\} \\
        Decoder Activation & \{ReLU, SeLU, GeLU, Swish\} \\
        Decoder Regularization (L2) & \{0.0, 0.001, 0.01\} \\
        Decoder Dropout & \{0.0, 0.1, 0.2, 0.3, 0.4, 0.5\} \\
        Decoder Batch Normalization & \{True\} 
        \down\\ \hline
\end{tabular}}{Hyperparameter search space in our tuning experiments, using Bayesian Optimization. The Units, Activation, Regularization, Dropout, and Batch Normalization explored values were the same for each layer, regardless of how many layers were chosen each time for both the encoder and decoder.}
\end{table}

\subsection{Basic Linear Autoencoder}

To establish a simple baseline, we also implement a \textit{linear autoencoder}, which can be seen as the most basic form of the autoencoder paradigm. The general structure is identical to that of the non-linear autoencoder: an encoder maps the input responses into a compressed latent space, and a decoder reconstructs the original input from this representation. The crucial difference is that we remove all non-linearities and intermediate layers. In practice, this means that both the encoder and decoder are reduced to single linear transformations:
\begin{equation*}
    \mathbf{z} = W_e \mathbf{x}, \qquad 
    \hat{\mathbf{x}} = W_d \mathbf{z},
\end{equation*}
where $W_e \in \mathbb{R}^{m \times d}$ is the encoder matrix projecting from the input dimension $d$ into a latent dimension $m$, and $W_d \in \mathbb{R}^{d \times m}$ is the decoder matrix projecting back into the original space.

\noindent
The linear autoencoder is restricted to capturing only linear correlations in the data. This results in a model that is mathematically simpler and computationally lighter, but also less expressive than its non-linear counterpart. Despite this limitation, the linear autoencoder provides a baseline against which the added value of deeper, non-linear architectures can be assessed. 
Linear autoencoders can be viewed as similar to PCA, where some constraints, such as additional orthogonality, have been omitted. 
% Second, it acts as a conceptual precursor to PCA, which can be viewed as a special case of a linear autoencoder with additional orthogonality constraints on the projection matrices. 

\noindent
In terms of training, we retain the same reconstruction-based objective as in the non-linear case. The model is optimized using the base loss function described in the previous section, ensuring comparability with more complex autoencoder architectures.

\subsection{Bayesian Probabilistic Network}

We model the joint distribution of the categorical survey variables with a tree-structured Bayesian network learned via the Chow–Liu algorithm~\citep{chow1968}. This family factors the joint as a product of local marginals and pairwise conditionals arranged on a tree, generating a fully-unsupervised scorer of respondent typicality. Inattentive responses, which fail to respect the dominant conditional dependencies among items, receive low likelihood under the learned model and are thus flagged as anomalies.

Let $\mathbf{X}=(X_1,\dots,X_d)$ be discrete variables, where $X_j \in \{1,\dots,K_j\}$ are the categories of question $j$ (we treat "No Answer" as an explicit category). A tree-structured Bayesian network $T$ consists of a root variable $X_r$ and directed edges from each parent to children, forming a directed acyclic graph with $d-1$ edges. The joint factorizes as
\begin{equation*}
    p_T(\mathbf{x})
    \;=\;
    p(x_r)\;\prod_{j\neq r} p\bigl(x_j \mid x_{\pi(j)}\bigr),
\end{equation*}
where $\pi(j)$ denotes the parent of node $j$ in the tree.

Given data $\mathcal{D}=\{\mathbf{x}^{(n)}\}_{n=1}^N$, the Chow–Liu algorithm finds the maximum-likelihood tree-structured approximation to the empirical distribution. It proceeds by (i) estimating the pairwise mutual information (MI) between all variable pairs and (ii) selecting the maximum-weight spanning tree under these MI weights. Concretely, for variables $X_i$ and $X_j$, we form the smoothed empirical joint
\begin{align*}
\widehat{p}_{ij}(a,b)
&= \frac{N_{ij}(a,b)+\alpha}{N + \alpha K_i K_j}, \\[6pt]
\widehat{p}_i(a)
&= \sum_{b}\widehat{p}_{ij}(a,b), \\[6pt]
\widehat{p}_j(b)
&= \sum_{a}\widehat{p}_{ij}(a,b).
\end{align*}

with Laplace smoothing $\alpha>0$ (we use $\alpha{=}1$). $N_{ij}(a,b)$ is the number of co-appearances of $X_i=a, X_j=b$. The (natural-log) mutual information is
\begin{equation*}
    \widehat{I}(X_i;X_j)
    \;=\;
    \sum_{a=1}^{K_i}\sum_{b=1}^{K_j}
    \widehat{p}_{ij}(a,b)\;
    \log\!\frac{\widehat{p}_{ij}(a,b)}{\widehat{p}_i(a)\,\widehat{p}_j(b)}.
\end{equation*}
Let $G$ be the complete graph on $\{1,\dots,d\}$ with edge weights $\widehat{I}(X_i;X_j)$. Any maximum-weight spanning tree $E^\star$ of $G$ defines the undirected structure; orienting its edges from a chosen root $r$ yields the directed tree. In practice we pick $r$ as the “MI-central” node (largest sum of MI to others) and the tree is obtained with a standard maximum-spanning-tree algorithm (Prim’s algorithm).

Given the directed tree, we estimate the root marginal and each conditional probability table (CPT) with Laplace smoothing. For the root:
\begin{equation*}
    \widehat{p}(x_r{=}k)
    \;=\;
    \frac{N_r(k)+\alpha}{N + \alpha K_r}
    \qquad (k=1,\dots,K_r).
\end{equation*}
For every child–parent pair $(c,p)$:
\begin{align*}
&\widehat{p}(x_c{=}k_c \mid x_p{=}k_p)
= \frac{N_{pc}(k_p,k_c)+\alpha}{N_{p}(k_p)+\alpha K_c}, \\[6pt]
&\quad k_p=1,\dots,K_p,\;\; k_c=1,\dots,K_c.
\end{align*}

Smoothing guards against zero counts, stabilizing both MI estimation and likelihoods in sparse, high-cardinality settings.

For a respondent $\mathbf{x}$, the (smoothed) log-likelihood under the learned tree is
\begin{equation*}
    \log \widehat{p}_T(\mathbf{x})
    \;=\;
    \log \widehat{p}(x_r)
    \;+\;
    \sum_{j\neq r} \log \widehat{p}\bigl(x_j \mid x_{\pi(j)}\bigr).
\end{equation*}

For ranking, we follow a rank-based percentile of typicality computed from the total log-likelihoods. Let $N$ be the number of respondents and let $r(\mathbf{x})\in\{1,\dots,N\}$ be the rank of $\mathbf{x}$ when sorting $\log \widehat{p}_T(\mathbf{x})$ in descending order (so $r{=}1$ is most typical, highest log-likelihood). We define
\[
\mathrm{pct}(\mathbf{x})
\;=\;
1 \;-\; \frac{r(\mathbf{x})-1}{\,N-1\,}
\;\in\;[0,1],
\]
so that $\mathrm{pct}{=}1$ denotes the most typical respondents and $\mathrm{pct}{=}0$ the least typical. Inattentiveness detection uses this statistic directly by ranking respondents in \emph{ascending} order of $\mathrm{pct}$. In other words, this ranking criterion is the equivalent of the reconstruction loss ranking in the Autoencoders case.

A Chow–Liu tree explicitly encodes dominant \emph{pairwise} dependencies across items (e.g., within batteries or between logically related questions). Attentive respondents tend to respect these dependencies and thus achieve high likelihood. Inattentive respondents, by contrast, break the learned conditional structure (e.g., incoherent or random choices), driving down their likelihoods. This leverages the same “structured patterns compress, noise does not” intuition as reconstruction-based methods, but from a probabilistic perspective with interpretable local dependencies.

% The Chow–Liu model is a special case of a discrete Bayesian network whose underlying directed graph is a tree. Unlike PCA or linear autoencoders, which reconstruct via linear projections in a Euclidean space, Chow–Liu models the probability of each joint configuration via discrete CPTs and scores atypicality with likelihood rather than squared error. It thus complements reconstruction-based methods with a generative, dependency-aware view of survey behavior.

A toy example of how this network works is provided in Appendix~\ref{appBayesian}.

%% file: sections/experiments.tex
\section{Experimental Results}\label{sec:Res}

\subsection{Metrics used for the Evaluation}

\subsubsection{Reconstruction Evaluation}
We first assess how accurately the model reconstructs the original data.  
For each variable, the accuracy, i.e., the proportion of correctly reconstructed values, is computed as
\[
\text{Accuracy}(v_i) = \frac{1}{N} \sum_{j=1}^{N} \mathbb{I}(x_{j,i} = \hat{x}_{j,i}),
\]
where $\mathbb{I}(\cdot)$ is the indicator function, and $x_{j,i}$ and $\hat{x}_{j,i}$ denote the true and reconstructed values of variable $v_i$ for sample $j$.  
The mean reconstruction accuracy across all $|\mathcal V|$ variables is
\[
\text{Mean Accuracy} = \frac{1}{|\mathcal V|} \sum_{i=1}^{|\mathcal V|} \text{Accuracy}(v_i).
\]
To benchmark performance, we compare this against a majority-class baseline for each variable:
\[
\text{Baseline Accuracy}(v_i) = \max_{c \in \mathcal{C}_i} \frac{1}{N} \sum_{j=1}^{N} \mathbb{I}(x_{j,i}=c),
\]
where $\mathcal{C}_i$ is the set of unique categorical values.  
The model’s improvement over the baseline is quantified by the Lift metric:
\[
\text{Lift} = \frac{1}{|\mathcal V|} \sum_{i=1}^{|\mathcal V|} \frac{\text{Accuracy}(v_i)}{\text{Baseline Accuracy}(v_i)},
\]
where Lift $>1$ indicates superior performance relative to majority-class prediction.  

To complement accuracy, we compute a One-vs-All ROC AUC metric, ORA, capturing how well the model ranks correct responses above incorrect ones.  
For variable $v_i$ with categories $\mathcal{C}_i$, 
\[
\text{ORA}(v_i) = \frac{1}{|\mathcal{C}_i|} \sum_{c \in \mathcal{C}_i} \text{ROC AUC}(c),
\]
and the overall score is 
\[
\text{Mean ORA} = \frac{1}{|\mathcal V|} \sum_{i=1}^{|\mathcal V|} \text{ORA}(v_i).
\]

\subsubsection{Randomness Detection Evaluation}
We evaluate the model’s ability to detect inattentive respondents, using attention-check outcomes as ground-truth labels. Because some datasets contain multiple checks, metrics are computed separately for each.

After computing a reconstruction error (or negative log-likelihood for Chow-Liu) for each respondent, we rank the respondents in descending order. This frames inattentiveness detection as an information-retrieval task: the goal is to place careless respondents at the top of the list.

\paragraph{Ranking Metrics:}
We report \textbf{Recall@h}, where $h$ is the true number of inattentive respondents in the dataset. This metric (also known as ``Success Rate'' in anomaly detection literature) measures the proportion of true anomalies captured if we set the rejection threshold exactly at the true contamination rate:
\[
\text{Recall@h} = \frac{|\{\text{inattentive users in top-}h\}|}{h}.
\]
We also report \textbf{Precision@k} for fixed operational cutoffs ($k \in \{10, 50, 100\}$) to simulate a ``human-in-the-loop'' auditing scenario where a researcher checks the top $k$ flagged cases:
\[
\text{Precision@k} = \frac{|\{\text{inattentive users in top-}k\}|}{k}.
\]
Note that Precision@$h$ = Recall@$h$.

To evaluate how well inattentive respondents appear near the top of the ranking, we compute the normalized discounted cumulative gain:
\[
\text{DCG@h} = \sum_{i=1}^{h} \frac{\mathbb{I}(\text{inattentive at rank } i)}{\log_2(i+1)}
\]
\[
\text{NDCG@h} = \frac{\text{DCG@h}}{\text{IDCG@h}},
\]
where IDCG@h represents the ideal (perfect) ranking.

\paragraph{Discriminative Metrics:}
Finally, we assess global separation performance independent of specific thresholds using the ROC curve, with reconstruction error as the decision threshold:
\[
\text{TPR} = \frac{\text{TP}}{\text{TP}+\text{FN}}, \qquad
\text{FPR} = \frac{\text{FP}}{\text{FP}+\text{TN}}.
\]
Plotting TPR versus FPR across thresholds yields the ROC curve, and the area under it
\[
\text{AUC} = \int_{0}^{1} \text{TPR}(\text{FPR}) \, d\text{FPR}
\]
quantifies ranking performance, where higher AUC values indicate stronger separation between attentive and inattentive respondents. To ensure our AUC estimates are robust to the ``train-on-test'' concern raised in anomaly detection, we confirmed these values against the out-of-sample scores generated by our cross-validation check. The reported AUCs therefore reflect the model's genuine ability to learn the structural difference between attentive and inattentive behaviors, rather than overfitting to specific respondent noise.

\subsection{Results}

\subsubsection{Reconstruction Performance}

Table~\ref{tab:reconstruction_perf} contains the results for the various methods on various models for the task of reconstruction. We include two variations of Non-Linear AE: (a) one with $p=100$, where it is trained as a standard AE, (b) a PL variation with $p=85$ that showed the most promising results. 
Overall, all three unsupervised models achieve strong reconstruction performance across datasets, consistently outperforming the majority-class baseline predictors. Accuracy values typically exceed 80\%, and Lift scores are well above 1.3 in nearly all cases, demonstrating that these models successfully capture the regularities of survey response patterns without supervision. Among them, linear autoencoders show the most stable and competitive performance across datasets, often reaching the highest accuracy and Lift values (e.g.,~\citealp{uhalt2020},~\citealp{mastroianni2022}). This suggests that linear mappings can capture much of the covariance structure present in survey data. The non-linear autoencoder performs comparably, and in several cases slightly better (e.g.,~\citealp{ogrady2019},~\citealp{alvarez2019}). 

The Percentile-Loss variant, by design, attains marginally lower reconstruction scores than the other approaches. Because it focuses learning on the lower-error subset of samples, it trades overall reconstruction accuracy for robustness and selective sensitivity, which is an expected and desirable behavior for inattentiveness detection (though detrimental to pure reconstruction metrics).  In general, these findings show that simple linear transformations are strong baselines for capturing dominant structure in categorical survey data, while non-linear and percentile-based models become advantageous when robustness or modeling subtle dependencies is required. Finally, we note that the Bayesian Networks are only leveraged for the second task of inattentive participants detection, since they could not be modified to execute the reconstruction task.

\begin{table}
\TABLE
% \scriptsize
{Reconstruction results across datasets and methods \label{tab:reconstruction_perf}}
{\begin{tabular}{@{}l@{\quad}c@{\quad}c@{\quad}c@{\quad}c@{}}
\textbf{Dataset \& Method} & \textbf{Accuracy} & \textbf{Baseline Acc} & \textbf{Lift} & \textbf{ORA} \\
\hline\up
\cite{robinson2014} &&&& \\
\quad Non-Linear Autoencoder ($p=100$) & 81.37 & 58.27 & 1.53 & \textbf{0.73} \\
\quad Non-Linear Autoencoder ($p=85$) & 80.83 & 58.27 & 1.52 & 0.70 \\
\quad Linear Autoencoder & \textbf{83.16} & 58.27 & \textbf{1.58} & 0.70 \\
\addlinespace
\cite{pennycook2020} &&&& \\
\quad Non-Linear Autoencoder ($p=100$)  & 86.84 & 66.22 & 1.38 & 0.83 \\
\quad Non-Linear Autoencoder ($p=85$) & 86.86 & 66.22 & 1.38 & 0.83 \\
\quad Linear Autoencoder & \textbf{89.18} & 66.22 & \textbf{1.44} & \textbf{0.86} \\
\addlinespace

\cite{alvarez2019} &&&& \\
\quad Non-Linear Autoencoder ($p=100$)  & \textbf{90.31} & 52.56 & 2.05 & \textbf{0.88} \\
\quad Non-Linear Autoencoder ($p=85$) & 89.69 & 52.56 & 2.02 & 0.86\\
\quad Linear Autoencoder & 90.07 & 52.56 & \textbf{2.07} & 0.85 \\
\addlinespace
\cite{uhalt2020} &&&& \\
\quad Non-Linear Autoencoder ($p=100$)  & 85.94 & 35.66 & 2.49 & \textbf{0.89} \\
\quad Non-Linear Autoencoder ($p=85$) & 81.31 & 35.66 & 2.35 & 0.83 \\
\quad Linear Autoencoder  & \textbf{91.44} & 35.66 & \textbf{2.64} & \textbf{0.89} \\
\addlinespace
\cite{ogrady2019} &&&& \\
\quad Non-Linear Autoencoder ($p=100$)  & \textbf{86.89} & 58.78 & \textbf{1.70} & \textbf{0.89} \\
\quad Non-Linear Autoencoder ($p=85$) & 84.30 & 58.78 & 1.64 & 0.85 \\
\quad Linear Autoencoder  & 79.80 & 58.78 & 1.68 & 0.86 \\
\addlinespace
\cite{buchanan2018} &&&& \\
\quad Non-Linear Autoencoder ($p=100$)   & \textbf{94.62} & 50.87 & 2.18 & \textbf{0.90} \\
\quad Non-Linear Autoencoder ($p=85$) & 91.89 & 50.87 & 2.12 & 0.75 \\
\quad Linear Autoencoder  & 92.80 & 50.87 & \textbf{2.19} & 0.84 \\
\addlinespace
\cite{moss2023} &&&& \\
\quad Non-Linear Autoencoder ($p=100$)   & 83.46 & 59.78 & 1.51 & 0.76 \\
\quad Non-Linear Autoencoder ($p=85$) & 82.73 & 59.78 & 1.49 & 0.75 \\
\quad Linear Autoencoder  & \textbf{88.59} & 59.78 & \textbf{1.66} & \textbf{0.82} \\
\addlinespace
\cite{mastroianni2022} &&&& \\
\quad Non-Linear Autoencoder ($p=100$)   & 77.77 & 61.64 & 1.26 & 0.68 \\
\quad Non-Linear Autoencoder ($p=85$) & 77.57 & 61.64 & 1.25 & 0.68 \\
\quad Linear Autoencoder  & \textbf{83.11} & 61.64 & \textbf{1.34} & \textbf{0.72} \\
\addlinespace
\cite{ivanov2021} &&&& \\
\quad Non-Linear Autoencoder ($p=100$)   & 76.15 & 45.32 & 1.86 & 0.78 \\
\quad Non-Linear Autoencoder ($p=85$) & 74.69 & 45.32 & 1.82 & 0.76\\
\quad Linear Autoencoder  & \textbf{80.24} & 45.32 & \textbf{2.00} & \textbf{0.84}\\
\down\\ \hline
\end{tabular}}{Evaluation of model performance across datasets when we try to “predict” an out-of-sample attribute value. Acc refers to the mean accuracy of the model in reconstructing the original data. Baseline Acc (Baseline Accuracy) represents the accuracy if we “guess” the majority class as the value for each attribute. Lift is computed as the ratio of Accuracy to Baseline Acc, indicating the improvement of our model over a naive predictor. ORA denotes the One-Vs-All ROC AUC metric, with 0.5 as the baseline performance.}
\end{table}

% linear autoencoder

\subsubsection{Randomness Detection Performance}

Tables~\ref{tab:randomness_det1},~\ref{tab:randomness_det2}, and~\ref{tab:randomness_det3} illustrate the results for the second task of Inattentive Respondents Detection. 
Across datasets, all models achieve meaningful performance in detecting inattentive respondents, confirming that unsupervised learning can reliably distinguish coherent from incoherent response patterns without supervision (there is a clear baseline of 0.5 for the AUC metric). Overall, Chow–Liu Bayesian network emerges as the most consistently strong performer, frequently obtaining the highest AUC and precision scores across diverse datasets and attention-check configurations. The Chow–Liu network excels in settings where dependencies are more localized and categorical, leveraging pairwise conditional probabilities to detect subtle violations of response consistency. These complementary results indicate that probabilistic formulations are effective in capturing typicality in survey data. The non-linear autoencoder also achieves competitive recall in some datasets (e.g.,~\cite{alvarez2019} and~\cite{ivanov2021}).

The PL autoencoder consistently offers balanced results, often improving precision and AUC over the base autoencoder, especially in datasets with noisier attention-check labels or heterogeneous content. This supports the intuition that down-weighting high-error samples during training yields a model less influenced by outliers and more sensitive to inattentiveness during evaluation. In contrast, the linear autoencoder generally underperforms, particularly in small or highly discrete datasets, highlighting the limits of purely linear mappings for inattentiveness detection even though they remain strong in reconstruction. The broad consistency across methods demonstrates that unsupervised modeling can serve as a powerful, domain-agnostic approach for identifying inattentive respondents in behavioral data.

\begin{table}
\TABLE
% \scriptsize
{Randomness Detection results across datasets and methods (Part I) \label{tab:randomness_det1}}
{\begin{tabular}{@{}l@{\quad}c@{\quad}c@{\quad}c@{\quad}c@{\quad}c@{\quad}c@{\quad}c@{}}
\textbf{Dataset \& Method} & \textbf{h} & \textbf{R@h} & \textbf{P@10} & \textbf{P@50} & \textbf{P@100} & \textbf{NDCG@h} & \textbf{AUC} \\
\hline\up
\cite{robinson2014} &&&& \\
\quad Non-Linear Autoencoder ($p=100$) & 230 & 0.07 & \textbf{0.20} & 0.08 & \textbf{0.09} & 0.08 & 0.71 \\
\quad Non-Linear Autoencoder ($p=85$)  & 230 & \textbf{0.10} & \textbf{0.20} & \textbf{0.12} & \textbf{0.09} & \textbf{0.09} & 0.74\\
\quad Linear Autoencoder & 230 & 0.03 & 0.00 & 0.04 & 0.03 & 0.03 & 0.51 \\
\quad Chow-Liu & 230 & 0.08 & 0.00 & 0.08 & \textbf{0.09} & 0.08 & \textbf{0.75} \\
\addlinespace
\cite{pennycook2020} &&&& \\

\quad Attention 1 &&&&& \\

\quad \quad Non-Linear Autoencoder ($p=100$) & 636 & 0.73 & \textbf{1.00} & 0.88 & 0.84 & 0.75 & 0.52 \\
\quad \quad Non-Linear Autoencoder ($p=85$) & 636 & 0.74 & \textbf{1.00} & 0.86 & 0.87 & 0.76 & 0.52 \\
\quad \quad Linear Autoencoder & 636 & \textbf{0.75} & 0.70 & 0.84 & 0.76 & 0.75 & \textbf{0.53}\\
\quad \quad Chow-Liu & 636 & \textbf{0.75} & \textbf{1.00} & \textbf{0.94} & \textbf{0.91} & \textbf{0.77} & \textbf{0.53} \\

\quad Attention 2 &&&&& \\

\quad \quad Non-Linear Autoencoder ($p=100$) & 236 & 0.30 & 0.50 & 0.30 & 0.34 & 0.32 & 0.52  \\
\quad \quad Non-Linear Autoencoder ($p=85$) & 236 & 0.31 & \textbf{0.60} & 0.44 & 0.37 & 0.35 & \textbf{0.53} \\
\quad \quad Linear Autoencoder & 236 & 0.31 & 0.20 & 0.36 & 0.33 & 0.31 & \textbf{0.53} \\
\quad \quad Chow-Liu & 236 & \textbf{0.33} & 0.50 & \textbf{0.50} & \textbf{0.39} & \textbf{0.36} & \textbf{0.53} \\

\quad Attention 3 &&&&& \\

\quad \quad Non-Linear Autoencoder ($p=100$) & 68 & 0.25 & 0.30 & 0.26 & 0.21 & 0.24 & 0.59 \\
\quad \quad Non-Linear Autoencoder ($p=85$) & 68 & 0.25 & \textbf{0.40} & 0.28 & 0.22 & 0.30 & 0.62 \\
\quad \quad Linear Autoencoder & 68 & 0.08 & 0.00 & 0.06 & 0.09 & 0.06 & 0.53 \\
\quad \quad Chow-Liu & 68 & \textbf{0.31} & 0.30 & \textbf{0.32} & \textbf{0.26} &\textbf{0.33} & \textbf{0.67} \\

\quad Attention 4 &&&&& \\

\quad \quad Non-Linear Autoencoder ($p=100$) & 172 & 0.24 & 0.40 & 0.36 & 0.29 & 0.27 & 0.50 \\
\quad \quad Non-Linear Autoencoder ($p=85$) & 172 & 0.26 & \textbf{0.70} & 0.40 & 0.36 & 0.32 & 0.51\\
\quad \quad Linear Autoencoder & 172 & 0.20 & 0.40 & 0.24 & 0.20 & 0.21 & 0.50 \\
\quad \quad Chow-Liu & 172 & \textbf{0.28} & 0.50 & \textbf{0.48} & \textbf{0.38} & \textbf{0.33} & \textbf{0.52} \\

\quad Union &&&&& \\

\quad \quad Non-Linear Autoencoder ($p=100$) & 686 & 0.79 & \textbf{1.00} & 0.96 & 0.91 & 0.81 & 0.53 \\
\quad \quad Non-Linear Autoencoder ($p=85$) & 686 & 0.80 & \textbf{1.00} & 0.92 & 0.91 & 0.81 & 0.53 \\
\quad \quad Linear Autoencoder & 686 & \textbf{0.81} & 0.80 & 0.94 & 0.86 & \textbf{0.82} & \textbf{0.55} \\
\quad \quad Chow-Liu & 686 & 0.80 & \textbf{1.00} & \textbf{0.98} & \textbf{0.95} & \textbf{0.82} & 0.53 \\

\quad Intersection &&&&& \\

\quad \quad Non-Linear Autoencoder ($p=100$) & 28 & 0.14 & 0.10 & \textbf{0.12} & 0.07 & 0.12 & 0.56 \\
\quad \quad Non-Linear Autoencoder ($p=85$) & 28 & 0.14 & \textbf{0.30} & \textbf{0.12} & 0.11 & \textbf{0.24} & 0.61 \\
\quad \quad Linear Autoencoder & 28 & 0.04 & 0.00 & 0.04 & 0.03 & 0.03 & 0.50 \\
\quad \quad Chow-Liu & 28 & \textbf{0.18} & 0.20 & \textbf{0.12} & \textbf{0.13} & 0.23 & \textbf{0.64} \\

\addlinespace

\cite{alvarez2019} &&&& \\
\quad Non-Linear Autoencoder ($p=100$) & 975 & \textbf{0.69} & \textbf{1.00} & \textbf{1.00} & \textbf{1.00} & 0.73 & \textbf{0.80} \\
\quad Non-Linear Autoencoder ($p=85$) & 975 & 0.68 & \textbf{1.00} & \textbf{1.00} & 0.99 & 0.72 & 0.77 \\
\quad Linear Autoencoder & 975 & 0.43 & 0.30 & 0.30 & 0.39 & 0.43 & 0.58 \\
\quad Chow-Liu & 975 & \textbf{0.69} & \textbf{1.00} & \textbf{1.00} & \textbf{1.00} & \textbf{0.74} & \textbf{0.80} \\

\down\\ \hline
\end{tabular}}{Evaluation of inattentiveness detection across datasets. h represents the number of inattentive users identified as ground truth. R@h denotes Recall@h. At perfect ranking, where all inattentive users have a higher error than all attentive, R@h = 1. P@k denotes Precision@k. A separate evaluation is provided for the datasets where we had more attention checks. Union means we consider a sample as inattentive only when it failed in one of the attention checks. Intersection means we consider a sample as inattentive only when it fails in all attention checks.
}
% \caption{Evaluation of inattentiveness detection across datasets. h represents the number of inattentive users identified as ground truth. R@h denotes Recall@h. At perfect ranking, where all inattentive users have a higher error than all attentive, R@h = 1. P@k denotes Precision@k. A separate evaluation is provided for the datasets where we had more attention checks. Union means we consider a sample as inattentive only when it failed in one of the attention checks. Intersection means we consider a sample as inattentive only when it fails in all attention checks.
% }
\end{table}

\begin{table}
\TABLE
{Randomness Detection results across datasets and methods (Part II) \label{tab:randomness_det2}}
{\begin{tabular}{@{}l@{\quad}c@{\quad}c@{\quad}c@{\quad}c@{\quad}c@{\quad}c@{\quad}c@{}}
\textbf{Dataset \& Method} & \textbf{h} & \textbf{R@h} & \textbf{P@10} & \textbf{P@50} & \textbf{P@100} & \textbf{NDCG@h} & \textbf{AUC} \\
\hline\up

\cite{uhalt2020} &&&& \\
\quad Non-Linear Autoencoder ($p=100$) & 6 & 0.17 & 0.10 & 0.04 & \textbf{0.05} & 0.15 & 0.67\\
\quad Non-Linear Autoencoder ($p=85$) & 6 & 0.33 & \textbf{0.30} & \textbf{0.10} & \textbf{0.05} & 0.34 & 0.82 \\
\quad Linear Autoencoder & 6 & 0.00 & 0.00 & 0.02 & 0.03 & 0.00 & 0.60 \\
\quad Chow-Liu & 6 & \textbf{0.50} & \textbf{0.30} & \textbf{0.10} & \textbf{0.05} & \textbf{0.39} & \textbf{0.88} \\

\addlinespace

\cite{ogrady2019} &&&& \\
\quad Non-Linear Autoencoder ($p=100$)  & 20 & 0.10 & 0.00 & 0.04 & 0.04 & 0.07 & 0.55 \\
\quad Non-Linear Autoencoder ($p=85$) & 20 & \textbf{0.20} & \textbf{0.10} & 0.08 & 0.06 & \textbf{0.15} & 0.55 \\
\quad Linear Autoencoder  & 20 & 0.05 & \textbf{0.10} & 0.08 & 0.10 & 0.05 & 0.63 \\
\quad Chow-Liu & 20 & 0.15 & \textbf{0.10} & \textbf{0.16} & \textbf{0.13} & 0.14 & \textbf{0.76} \\

\addlinespace
\cite{buchanan2018} &&&& \\
\quad Non-Linear Autoencoder ($p=100$)   & 59 & 0.07 & 0.10 & 0.08 & 0.08 & 0.08 & 0.61 \\
\quad Non-Linear Autoencoder ($p=85$) & 59 & \textbf{0.27} & \textbf{0.80} & \textbf{0.28} & \textbf{0.21} & \textbf{0.34} & \textbf{0.72} \\
\quad Linear Autoencoder  & 59 & 0.08 & 0.10 & 0.06 & 0.07 & 0.07 & 0.51 \\
\quad Chow-Liu & 59 & 0.22 & 0.10 & 0.20 & 0.19 & 0.19 & 0.65 \\

\addlinespace
\cite{moss2023} &&&& \\
\quad Attention 1 &&&&& \\

\quad \quad Non-Linear Autoencoder ($p=100$) & 161 & 0.16 & 0.10 & 0.16 & 0.14 & 0.15 & 0.53 \\
\quad \quad Non-Linear Autoencoder ($p=85$) & 161 & \textbf{0.18} & \textbf{0.20} & 0.20 & \textbf{0.17} & \textbf{0.20} & \textbf{0.59} \\
\quad \quad Linear Autoencoder & 161 & 0.11 & \textbf{0.20} & 0.18 & 0.11 & 0.13 & 0.56 \\
\quad \quad Chow-Liu & 161 & 0.12 & \textbf{0.20} & \textbf{0.22} & 0.13 & 0.16 & 0.58 \\

\quad Attention 2 &&&&& \\

\quad \quad Non-Linear Autoencoder ($p=100$) & 140 & 0.11 & \textbf{0.30} & 0.20 & 0.13 & 0.12 & 0.54 \\
\quad \quad Non-Linear Autoencoder ($p=85$) & 140 & \textbf{0.21} & \textbf{0.30} & \textbf{0.24} & \textbf{0.22} & \textbf{0.24} & \textbf{0.60} \\
\quad \quad Linear Autoencoder & 140 & 0.06 & 0.20 & 0.08 & 0.06 & 0.08 & 0.54 \\
\quad \quad Chow-Liu & 140 & 0.11 & 0.20 & 0.14 & 0.12 & 0.15 & 0.59 \\

\quad Union &&&&& \\

\quad \quad Non-Linear Autoencoder ($p=100$) & 248 & 0.16 & \textbf{0.30} & 0.28 & 0.22 & 0.17 & 0.54 \\
\quad \quad Non-Linear Autoencoder ($p=85$) & 248 & 0.20 & \textbf{0.30} & \textbf{0.30} & \textbf{0.25} & \textbf{0.23} & \textbf{0.59} \\
\quad \quad Linear Autoencoder & 248 & 0.16 & \textbf{0.30} & 0.20 & 0.13 & 0.17 & 0.53 \\
\quad \quad Chow-Liu & 248 & \textbf{0.21} & \textbf{0.30} & 0.28 & 0.20 & \textbf{0.23} & 0.57\\

\quad Intersection &&&&& \\

\quad \quad Non-Linear Autoencoder ($p=100$) & 53 & 0.08 & 0.00 & 0.08 & 0.06 & 0.07 & 0.56 \\
\quad \quad Non-Linear Autoencoder ($p=85$) & 53 & \textbf{0.11} & \textbf{0.20} & \textbf{0.10} & \textbf{0.13} & \textbf{0.13} & \textbf{0.65} \\
\quad \quad Linear Autoencoder & 53 & 0.06 & 0.10 & 0.06 & 0.04 & 0.08 & 0.62 \\
\quad \quad Chow-Liu & 53 & 0.08 & 0.10 & 0.08 & 0.05 & \textbf{0.13} & 0.61 \\

\down\\ \hline
\end{tabular}}{Evaluation of inattentiveness detection across datasets. h represents the number of inattentive users identified as ground truth. R@h denotes Recall@h. At perfect ranking, where all inattentive users have a higher error than all attentive, R@h = 1. P@k denotes Precision@k. A separate evaluation is provided for the datasets where we had more attention checks. Union means we consider a sample as inattentive only when it failed in one of the attention checks. Intersection means we consider a sample as inattentive only when it fails in all attention checks.
}
\end{table}

\begin{table}
\TABLE
{Randomness Detection results across datasets and methods (Part III) \label{tab:randomness_det3}}
{\begin{tabular}{@{}l@{\quad}c@{\quad}c@{\quad}c@{\quad}c@{\quad}c@{\quad}c@{\quad}c@{}}
\textbf{Dataset \& Method} & \textbf{h} & \textbf{R@h} & \textbf{P@10} & \textbf{P@50} & \textbf{P@100} & \textbf{NDCG@h} & \textbf{AUC} \\
\hline\up

\cite{mastroianni2022} &&&& \\
\quad Non-Linear Autoencoder ($p=100$)   & 60 & \textbf{0.50} & \textbf{1.00} & \textbf{0.56} & 0.30 & \textbf{0.62} & 0.66 \\
\quad Non-Linear Autoencoder ($p=85$) & 60 & 0.43 & \textbf{1.00} & 0.52 & \textbf{0.31} & 0.57 & \textbf{0.68} \\
\quad Linear Autoencoder  & 60 & 0.05 & 0.00 & 0.04 & 0.04 & 0.04 & 0.52 \\
\quad Chow-Liu & 60 & 0.45 & \textbf{1.00} & 0.54 & 0.28 & 0.58 & 0.61 \\

\addlinespace

\cite{ivanov2021} &&&& \\
\quad Attention 1 &&&&& \\

\quad \quad Non-Linear Autoencoder ($p=100$) & 165 & 0.32 & 0.70 & 0.42 & 0.37 & 0.37 & 0.64 \\
\quad \quad Non-Linear Autoencoder ($p=85$) & 165 & 0.39 & \textbf{0.90} & \textbf{0.60} & 0.49 & 0.46 & 0.69 \\
\quad \quad Linear Autoencoder & 165 & 0.22 & 0.40 & 0.28 & 0.26 & 0.23 & 0.51 \\
\quad \quad Chow-Liu & 165 & \textbf{0.46} & \textbf{0.90} & 0.58 & \textbf{0.53} & \textbf{0.51} & \textbf{0.73} \\

\quad Attention 2 &&&&& \\

\quad \quad Non-Linear Autoencoder ($p=100$) & 55 & 0.09 & 0.20 & 0.10 & 0.13 & 0.15 & 0.61 \\
\quad \quad Non-Linear Autoencoder ($p=85$) & 55 & \textbf{0.22} & \textbf{0.30} & \textbf{0.20} & 0.21 & \textbf{0.20} & 0.69 \\
\quad \quad Linear Autoencoder & 55 & 0.09 & \textbf{0.30} & 0.10 & 0.07 & 0.13 & 0.51 \\
\quad \quad Chow-Liu & 55 & 0.18 & 0.10 & 0.16 & \textbf{0.23} & 0.16 & \textbf{0.76} \\

\quad Union &&&&& \\

\quad \quad Non-Linear Autoencoder ($p=100$) & 173 & 0.35 & 0.70 & 0.46 & 0.41 & 0.40 & 0.65 \\
\quad \quad Non-Linear Autoencoder ($p=85$) & 173 & 0.42 & \textbf{0.90} & \textbf{0.64} & 0.53 & 0.48 & 0.70 \\
\quad \quad Linear Autoencoder & 173 & 0.22 & 0.40 & 0.30 & 0.27 & 0.24 & 0.50 \\
\quad \quad Chow-Liu &  173 & \textbf{0.48} & \textbf{0.90} & 0.62 & \textbf{0.56} & \textbf{0.53} & \textbf{0.74}\\
\quad Intersection &&&&& \\

\quad \quad Non-Linear Autoencoder ($p=100$) & 47 & 0.06 & 0.20 & 0.06 & 0.09 & 0.13 & 0.59 \\
\quad \quad Non-Linear Autoencoder ($p=85$) & 47 & \textbf{0.15} & \textbf{0.30} & \textbf{0.16} & 0.17 & \textbf{0.15} & 0.67 \\
\quad \quad Linear Autoencoder & 47 & 0.09 & \textbf{0.30} & 0.08 & 0.06 & 0.12 & 0.52 \\
\quad \quad Chow-Liu & 47 & 0.09 & 0.10 & 0.12 & \textbf{0.20} & 0.08 & \textbf{0.76} \\
\down\\ \hline
\end{tabular}}{Evaluation of inattentiveness detection across datasets. h represents the number of inattentive users identified as ground truth. R@h denotes Recall@h. At perfect ranking, where all inattentive users have a higher error than all attentive, R@h = 1. P@k denotes Precision@k. A separate evaluation is provided for the datasets where we had more attention checks. Union means we consider a sample as inattentive only when it failed in one of the attention checks. Intersection means we consider a sample as random only when it fails in all attention checks.
}
\end{table}

%% file: sections/analysis.tex
\section{Analysis}

\subsection{When do Unsupervised Methods work well?}

All unsupervised models demonstrate meaningful ability to detect inattentive respondents, yet their performance varies across datasets and metrics. Understanding the sources of this variability is critical for assessing robustness and for guiding practical deployment in behavioral research. We therefore examine when and why these methods succeed or fail to capture inattentiveness effectively.  

We begin with a dataset-level analysis that tests whether detection variability relates to dataset characteristics: sample size (\textit{Samples}), number of variables (\textit{Variables}), total number of one-hot features (\textit{Features}), and average features per variable (\textit{AFV}). For the randomness-detection task, each dataset is summarized by the mean AUC across methods. For reconstruction, we use the mean Lift. We compute Pearson $r$ and Spearman $\rho$ correlations (two-sided tests; $n=9$ datasets) to capture linear and monotone associations, respectively.  

Table~\ref{tab:dataset_correlations} shows that dataset size and dimensionality (\textit{Samples}, \textit{Variables}, \textit{Features}, \textit{AFV}) are not reliably associated with detection performance (Pearson $|r|\!\le\!0.54$; Spearman $|{\rho}|\!\le\!0.48$; all $p{>}0.18$). In contrast, the mean reconstruction Lift correlates moderately with mean AUC ($r{\approx}0.69$, $p{\approx}0.04$; $\rho{\approx}0.68$), suggesting that when models better capture structured regularities, inattentiveness becomes more separable.\footnote{Estimates are based on the nine datasets in Table~\ref{tab:data_stats}. Method-specific correlations appear in Appendix~\ref{appendix_more_results}.}  

At the dataset level, these findings imply that scale and sparsity are not the dominant determinants of detection success. Instead, surveys whose dependency structure is well captured (i.e., higher reconstruction Lift) enable clearer differentiation between attentive and inattentive respondents. In practice, this means that the informativeness of survey structure, and not its size, governs how well unsupervised models detect inattentiveness.

\begin{table}[t]
\TABLE
{Correlations between dataset characteristics and performance  \label{tab:dataset_correlations}}
{\begin{tabular}{lcccc}
\toprule
 & \multicolumn{2}{c}{\textbf{Mean AUC}} & \multicolumn{2}{c}{\textbf{Mean Lift}}\\
\cmidrule(lr){2-3} \cmidrule(lr){4-5}
\textbf{Predictor} & Pearson $r$ & Spearman $\rho$ & Pearson $r$ & Spearman $\rho$\\
\midrule
Samples   & $0.20$ & $0.05$ & $-0.24$ & $-0.20$\\
Variables & $-0.46$ & $-0.16$ & $-0.47$ & $-0.46$ \\
Features  & $-0.37$ & $-0.21$ & $-0.54$ & $-0.48$ \\
AFV       & $0.11$  & $0.06$ & $0.10$ & $0.15$ \\
\addlinespace
\textbf{Mean Lift}  & $0.69^*$  & $0.68^*$ &  & \\
\bottomrule
\end{tabular}}
{Pearson and Spearman correlations (two‑sided) across $n{=}9$ datasets. Mean Lift $\leftrightarrow$ Mean AUC shows a moderate positive association ($p{\approx}0.04$), whereas other predictors show no reliable relationship to both reconstruction and randomness performance.}
\end{table}

How well unsupervised models distinguish attentive from inattentive respondents depends primarily on two factors: the internal structure of the survey and the quality of its attention-check labels. Datasets such as \cite{uhalt2020} and \cite{alvarez2019} represent best-case scenarios; they employ well-structured, multi-item scales in which several questions measure the same latent construct, and their attention checks directly test instruction compliance. This combination provides consistent response patterns that models can easily learn. By contrast, datasets like \cite{mastroianni2022} and \cite{buchanan2018} mix attitudinal, behavioral, and metadata-based variables, weakening cross-item regularities. Non linear models can still identify structure, but these patterns are less conceptually coherent, reducing separability between attentive and inattentive respondents.

The studies by \cite{pennycook2020} and \cite{moss2023} further illustrate how strongly detection performance depends on labeling precision. Both include repeated thematic items that should, in principle, expose clear patterns among attentive respondents, yet their attention checks differ in diagnosticity. Self-reports and comprehension checks yield noisy, heterogeneous labels, whereas strict instruction-following items produce far clearer ground truth. When inattentiveness is defined loosely (e.g., failing any check), model performance appears weaker; when only the most diagnostic or “fail-all” criteria are used, separation improves markedly (e.g., “Attention 3” in \cite{pennycook2020} achieves AUC = 0.67).  

Overall, these findings show that coherent survey design provides models with meaningful structure to learn, but precise labeling determines how fully that potential is realized. The challenge of defining inattentiveness, especially when attention checks are themselves imperfect, remains a limitation of our evaluation and, more broadly, of research on response quality.

\subsection{Reconstruction–Detection Trade-off}\label{subsec:pl_tradeoff}

Standard AE minimize mean reconstruction loss, learning to reproduce all inputs. For inattentiveness detection, however, the objective is to model the typical manifold while allowing incoherent responses to remain poorly reconstructed. The PL objective achieves this by averaging the lowest-error $p$\% of samples in each batch: smaller $p$ values down-weight high-loss cases during training, leaving them anomalous at test time. This creates a trade-off: higher $p$ improves reconstruction performance, whereas lower $p$ enhances anomaly sensitivity.

\begin{figure}[t]
  \centering
  \includegraphics[width=0.62\linewidth]{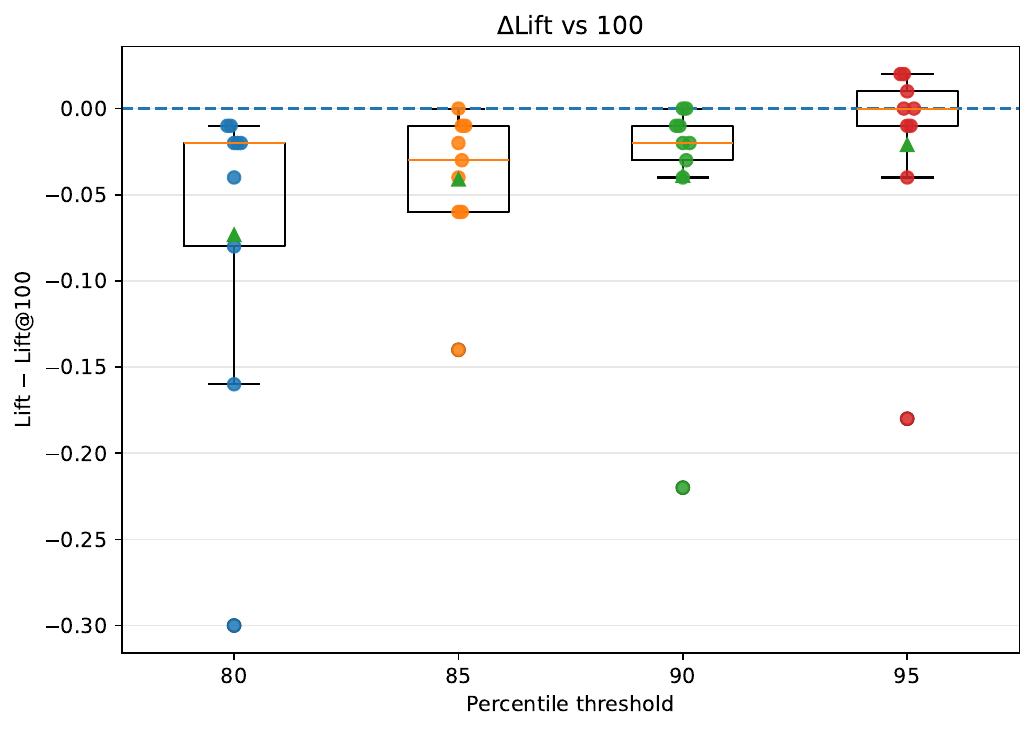}
  \caption{Percentile Loss trade-off: $\Delta \text{Lift}$ (reconstruction) vs.\ $p$ relative to $p{=}100$. Medians are negative at $p{=}80/85/90$ and approach $0$ at $p{=}95$, indicating reconstruction improves monotonically as $p\!\to\!100$.}
  \label{fig:pl_box_lift}
\end{figure}

\begin{figure}[t]
  \centering
  \includegraphics[width=0.62\linewidth]{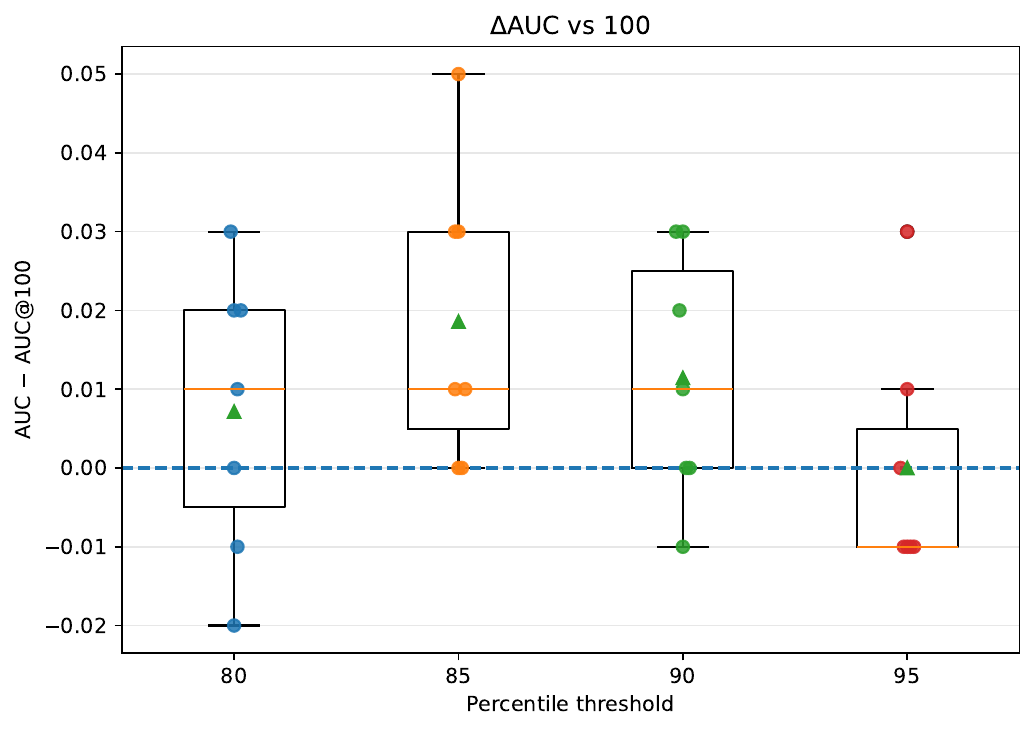}
  \caption{Percentile Loss trade-off: $\Delta \text{AUC}$ (randomness detection) vs.\ $p$ relative to $p{=}100$. Gains at $p{=}80/85/90$ and attenuation at $p{=}95$ suggest a broad optimum near $p\approx 85\text{--}90$.}
  \label{fig:pl_box_auc}
\end{figure}

Across datasets, we visualize changes relative to the full-batch objective ($p{=}100$) in Figs.~\ref{fig:pl_box_lift}–\ref{fig:pl_box_auc}. As $p$ increases, $\Delta \text{Lift}$ rises monotonically toward zero (better reconstruction), while $\Delta \text{AUC}$ follows an inverted-U pattern with gains at $p{=}80$–$90$ and attenuation by $p{=}95$. Paired within-dataset tests confirm this pattern: detection AUC improves most reliably at $p{=}85$–$90$ ($\overline{\Delta}{\approx}{+}0.04$, $p{\approx}0.001$), with smaller gains at $p{=}80$ and none by $p{=}95$; reconstruction performance, by contrast, steadily increases as $p{\to}100$ (mean $\overline{\Delta}{\approx}{-}0.07$ at $p{=}80$). 

Overall, reconstruction improves monotonically with $p$, while detection peaks broadly around $p{\approx}85$–$90$. At these values, the model retains enough typical cases to learn the main manifold while keeping high-loss (incoherent) responses distinct. As $p{\to}100$, overfitting to all inputs erodes anomaly contrast; if $p$ is too low, the model over-focuses on trivial patterns, reducing generalization. This can be generalized as a reconstruction-randomness detection tradeoff: improving the model’s ability to reconstruct all inputs inevitably reduces its ability to detect atypical ones.

%% file: sections/implications.tex
\section{Practical Implications: The Economics of Automated Quality Control}\label{sec:Implications}

Our findings demonstrate that unsupervised models can reliably detect inattentiveness, particularly in well-structured surveys. However, for survey platforms and researchers, the decision to adopt this framework is not driven solely by detection accuracy (AUC), but by the economic trade-offs between data quality, respondent burden, and operational cost. In this section, we formalize this decision problem and derive the conditions under which unsupervised screening dominates traditional attention checks.

\subsection{A Cost-Benefit Model of Screening}
We model the screening process as a cost-minimization problem. Let $N$ be the total number of respondents. Traditional quality control relies on embedded attention checks (AC), while our proposed method relies on Unsupervised Modeling (UM). The total cost function $L$ for each approach can be decomposed into \textit{Burden Cost} (the cost imposed on respondents) and \textit{Error Cost} (the cost of retained noise or discarded valid data).

\textbf{1. The Cost of Attention Checks ($L_{AC}$).} 
Attention checks impose a cognitive tax and increase survey duration. Let $c_{tax}$ be the marginal cost per respondent of including checks (monetized time, increased dropout risk, or measurement reactivity). If checks are treated as "ground truth," we assume their error rate is negligible, but the burden applies to all $N$ respondents:
\begin{equation}
    L_{AC} = N \cdot c_{tax}
\end{equation}

\textbf{2. The Cost of Unsupervised Modeling ($L_{UM}$).}
Our label-free approach has zero respondent burden ($c_{tax} = 0$). However, because it is a probabilistic proxy, it introduces classification errors. Let $\alpha$ be the prevalence of inattentive respondents. Let $FNR$ be the False Negative Rate (inattentive respondents missed) and $FPR$ be the False Positive Rate (attentive respondents flagged).
Let $c_{noise}$ be the cost of including an inattentive respondent in the final dataset (reduced statistical power, potential for spurious correlations).
Let $c_{discard}$ be the cost of falsely excluding a valid respondent (wasted recruitment fees, reduced sample size).
The total cost for the unsupervised approach is:
\begin{align}
    L_{UM} & =  C_{compute} + N \cdot C_{model} \\
    C_{model} & =  \alpha \cdot FNR \cdot c_{noise} + (1-\alpha) \cdot FPR \cdot c_{discard}
\end{align}
where $C_{compute}$ is the fixed cost of training the model, which is negligible for modern hardware ($\approx 0$ per respondent).

\textbf{3. The Decision Rule.}
A platform should switch from Attention Checks to Unsupervised Modeling when $L_{UM} < L_{AC}$. Assuming $C_{compute} \approx 0$, the condition becomes:
\begin{equation}
    \alpha \cdot FNR \cdot c_{noise} + (1-\alpha) \cdot FPR \cdot c_{discard} < c_{tax}
\end{equation}
This inequality reveals the precise value proposition of our framework. In settings where respondent time is expensive (e.g., expert surveys where $c_{tax}$ is high) or where "measurement reactivity" threatens validity, the unsupervised approach is economically superior \textit{even if it is less accurate than explicit checks}, provided the error rates ($FNR, FPR$) remain bounded. Conversely, in low-stakes crowdsourcing where $c_{tax}$ is low but data integrity ($c_{noise}$) is paramount, a hybrid approach may be optimal.

\subsection{Operationalizing the Threshold}
A critical managerial decision is setting the sensitivity of the detector. Our analysis of the Percentile Loss (PL) hyperparameter $p$ (Section 6.2) provides a direct lever for this trade-off.
\begin{itemize}
    \item \textbf{Minimizing Noise ($c_{noise} \gg c_{discard}$):} If the cost of bad data is high (e.g., medical or policy research), practitioners should target the ``robust region'' we identified ($p \in [85, 90]$). As shown in Figure 4, this range maximizes the separation between the typical manifold and anomalies, effectively lowering $FNR$ at the expense of slightly higher reconstruction error.
    \item \textbf{Minimizing Discard ($c_{discard} \gg c_{noise}$):} If recruitment is difficult (e.g., rare populations), researchers should adopt a conservative exclusion strategy. Rather than applying a fixed threshold, we recommend ranking respondents by reconstruction error and inspecting the "elbow" of the error distribution.
\end{itemize}

\subsection{Design as Governance}
Perhaps our most actionable finding for Information Systems is that detection is endogenous to design. We showed that surveys with coherent, overlapping item batteries (high AFV or Mean Lift) facilitate significantly better separation.
This implies that \textit{survey design} is a governance mechanism. Platforms seeking to automate quality control need not rely solely on better algorithms; they can instead enforce design guidelines—specifically, requiring valid scale construction with redundant indicators. This increases the "signal-to-noise" ratio of the response data, effectively driving down both $FNR$ and $FPR$ in Equation (3) without adding explicit check items.

\subsection{Ethical Deployment: Human-in-the-Loop}
Finally, we caution against fully automated exclusion based solely on unsupervised scores. As noted in Section 4.3, autoencoders flag "incoherence," which can sometimes overlap with legitimate but minority perspectives. To mitigate fairness risks ($c_{discard}$), we recommend a two-stage governance process:
\begin{enumerate}
    \item \textbf{Automated Flagging:} Use the reconstruction error score to rank the top $q\%$ of respondents as "High Risk."
    \item \textbf{Human Review:} Present these high-risk profiles to a human auditor. The visualized reconstruction errors (as in Figure 1) serve as an explainable interface, highlighting exactly \textit{which} items deviated from the expected pattern.
\end{enumerate}
This "Augmented Intelligence" approach balances the scalability of unsupervised learning with the nuance of human judgment, ensuring that data quality efforts do not inadvertently enforce homogeneity.

%% file: sections/conclusions.tex
\section{Conclusion}\label{sec:Conclusion}

This paper advances a label–free, survey–agnostic framework for inattentiveness detection that combines complementary modeling views: geometric reconstruction (AE) and probabilistic dependency modeling (Chow–Liu). Across nine heterogeneous, real–world datasets, we show that modeling cross–item regularities suffices to surface incoherent respondents without supervision, yielding a practical two–stage workflow; unsupervised modeling followed by respondent ranking via reconstruction error or (negative) likelihood. 
Methodologically, we adapt autoencoders to categorical survey data with variable–level loss weighting and introduce Percentile Loss, documenting a clear reconstruction–detection trade–off with a robust operating region at $p\!\in[85,90]$ for inattentiveness detection. 
Substantively, we show that instruments with coherent, overlapping item batteries produce stable covariance and dependency patterns that our models leverage to separate attentive from inattentive respondents more effectively, linking psychometric design choices directly to unsupervised anomaly detection performance. 
Together, these results convert inattentiveness screening from ad hoc rules and reactive checks into a scalable, transparent diagnostic that generalize across surveys, languages, and populations.

%% file: sections/limitations.tex
\section{Limitations and Future Research}
\paragraph{Evaluation labels.} Our detection performance is evaluated against attention–check outcomes that are known proxies of inattentiveness and may be noisy. While our findings are robust across multiple datasets and checks, future work should incorporate alternative ground truths (e.g. eye-tracking inattentiveness labels) to quantify sensitivity to label noise more precisely.

\paragraph{Text restrictions and modality coverage.} We focus exclusively on structured categorical variables and discretized numeric responses, intentionally excluding open–ended text fields to maintain comparability across datasets. This design choice limits generalization to instruments with substantial free–text content. A natural extension is to integrate text embeddings (e.g., sentence–level representations) into our reconstruction and likelihood pipelines, enabling unified multimodal screening of mixed–type survey data.

\paragraph{Modeling scope.} We evaluate linear/nonlinear autoencoders (with PL), and a Chow–Liu Bayesian network. Other probabilistic families (e.g., mixture–IRT with response times, factor–mixture models) and modern tabular generative models (e.g., flows, diffusion on categorical spaces) may further improve typicality scoring.

\paragraph{Downstream impact on scientific conclusions.} Ultimately, we care whether the removal of inattentive users affects the conclusions of the study that uses the results of the survey. To analyse whether detecting inattentive users is useful, we need to replicate the results of all nine studies and examine how the results of each paper vary with different degrees of inattentiveness removal. This process would be similar to \citep{cimpian2018bias}, which explores how the estimates from census data change when we remove mischievous responders. We note, though, that such an analysis, which requires replicating work from nine papers, is a very serious undertaking, and we leave this as part of future work.

%% file: sections/appendices.tex
\begin{APPENDICES}
\section{Bayesian Networks: An Example} \label{appBayesian}

Suppose we have a dataset that consists of 5 variables: age, sex, grade, height, and weight. 
We discretize each variable into a few intuitive categories:
\begin{itemize}
    \item \(\texttt{age} \in \{A_1\!=\!12\text{--}14,\; A_2\!=\!15\text{--}17\}\) 
    \item \(\texttt{sex} \in \{M,F\}\)
    \item \(\texttt{grade} \in \{G_1=\text{middle},\; G_2=\text{high}\}\)
    \item \(\texttt{height} \in \{H_1=\text{short},\; H_2=\text{avg},\; H_3=\text{tall}\}\)
    \item \(\texttt{weight} \in \{W_1=\text{light},\; W_2=\text{avg},\; W_3=\text{heavy}\}\)
\end{itemize}
Intuitive real-world dependencies we wish to capture: 
\(\texttt{height}\leftrightarrow\texttt{weight}\) (strong), 
\(\texttt{age}\to\texttt{grade}\) (strong; older students are likelier in high school),
\(\texttt{height}\leftrightarrow\texttt{age}\) (moderate; older tend to be taller),
\(\texttt{height}\leftrightarrow\texttt{sex}\) (moderate; males are slightly taller on average).

We assume a tiny dataset (\(N{=}10\)) whose sufficient statistics reflect the above:
\begin{align*}
\text{Root (height):}\quad
& \#(H_1,H_2,H_3)=(2,5,3).\\[2pt]
\text{Weight}\mid\text{height}:\quad
& \#(W_1,W_2,W_3\mid H_1)=(2,0,0),\\
&  \#(W_1,W_2,W_3\mid H_2)=(1,3,1),\\
& \#(W_1,W_2,W_3\mid H_3)=(0,1,2).\smallskip\\
\text{Sex}\mid\text{height}:\quad
& \#(M,F\mid H_1)=(0,2),\\
&  \#(M,F\mid H_2)=(3,2),\\
&  \#(M,F\mid H_3)=(2,1).\\[2pt]
\text{Age}\mid\text{height}:\quad
& \#(A_1,A_2\mid H_1)=(2,0),\\
&  \#(A_1,A_2\mid H_2)=(2,3),\\
&  \#(A_1,A_2\mid H_3)=(0,3).\\[2pt]
\text{Grade}\mid\text{age}:\quad
& \#(G_1,G_2\mid A_1)=(4,0),\\
&  \#(G_1,G_2\mid A_2)=(1,5).
\end{align*}

We use Laplace smoothing \(\alpha=1\) in all estimates.
Computing pairwise MI on the empirical (smoothed) joints, the largest weights arise on 
\((\texttt{height},\texttt{weight})\) and \((\texttt{age},\texttt{grade})\).
As the maximum spanning tree must connect all variables, a natural ML tree selects
\(\texttt{height}\) as a root (high MI centrality), and adds edges 
\(\texttt{height}\to\texttt{weight}\), \(\texttt{height}\to\texttt{sex}\), \(\texttt{height}\to\texttt{age}\), and finally \(\texttt{age}\to\texttt{grade}\).

\begin{figure}[t]
\centering
\begin{tikzpicture}[>=stealth, node distance=2.2cm,
    var/.style={draw, rounded corners, minimum width=18mm, minimum height=6mm, align=center}]
\node[var] (H) {height};
\node[var, below left=1.2cm and 1.2cm of H] (W) {weight};
\node[var, below right=1.2cm and 1.2cm of H] (S) {sex};
\node[var, right=2.8cm of H] (A) {age};
\node[var, right=2.8cm of A] (G) {grade};

\draw[->, thick] (H) -- (W) node[midway, left] {\small strong};
\draw[->, thick] (H) -- (S) node[midway, right] {\small moderate};
\draw[->, thick] (H) -- (A) node[midway, above] {\small moderate};
\draw[->, thick] (A) -- (G) node[midway, above] {\small strong};
\end{tikzpicture}
\caption{Toy Chow--Liu tree: \texttt{height} as root; edges reflect strongest mutual information.}
\end{figure}
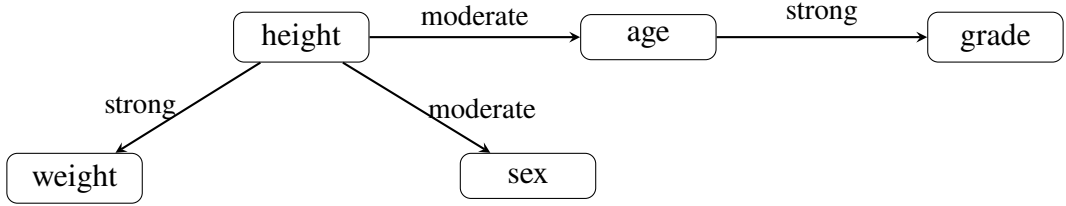

With \(\alpha{=}1\) and category sizes \(K_H{=}3\), \(K_W{=}3\), \(K_S{=}2\), \(K_A{=}2\), \(K_G{=}2\):
\[
\widehat{p}(H)=\frac{\#(H)+\alpha}{N+\alpha K_H}
=\Bigl(\tfrac{3}{13},\tfrac{6}{13},\tfrac{4}{13}\Bigr)
= (H_1,H_2,H_3).
\]
Conditionals (row-normalized with Laplace):
\[
\setlength{\arraycolsep}{0.6em}
\begin{aligned}
&\begin{array}{@{}l l@{}}
\widehat{p}(W\mid H_1)=\bigl(\tfrac{3}{5},\,\tfrac{1}{5},\,\tfrac{1}{5}\bigr) &
\widehat{p}(W\mid H_2)=\bigl(\tfrac{2}{8},\,\tfrac{4}{8},\,\tfrac{2}{8}\bigr) \\[0.5cm]
\multicolumn{2}{c}{\widehat{p}(W\mid H_3)=\bigl(\tfrac{1}{6},\,\tfrac{2}{6},\,\tfrac{3}{6}\bigr)}
\end{array} \\[0.5cm]
&\begin{array}{@{}l l@{}}
\widehat{p}(S\mid H_1)=\bigl(\tfrac{1}{4},\,\tfrac{3}{4}\bigr) &
\widehat{p}(S\mid H_2)=\bigl(\tfrac{4}{7},\,\tfrac{3}{7}\bigr) \\[0.5cm]
\multicolumn{2}{c}{\widehat{p}(S\mid H_3)=\bigl(\tfrac{3}{5},\,\tfrac{2}{5}\bigr)}
\end{array} \\[0.5cm]
&\begin{array}{@{}l l@{}}
\widehat{p}(A\mid H_1)=\bigl(\tfrac{3}{4},\,\tfrac{1}{4}\bigr) &
\widehat{p}(A\mid H_2)=\bigl(\tfrac{3}{7},\,\tfrac{4}{7}\bigr) \\[0.5cm]
\multicolumn{2}{c}{\widehat{p}(A\mid H_3)=\bigl(\tfrac{1}{5},\,\tfrac{4}{5}\bigr)}
\end{array} \\[0.5cm]
&\begin{array}{@{}l l@{}}
\widehat{p}(G\mid A_1)=\bigl(\tfrac{5}{6},\,\tfrac{1}{6}\bigr) &
\widehat{p}(G\mid A_2)=\bigl(\tfrac{1}{4},\,\tfrac{3}{4}\bigr)
\end{array}
\end{aligned}
\]

How the tree would be created:
\begin{enumerate}
    \item Compute smoothed pairwise joints \(\widehat{p}_{ij}\) and MIs \(\widehat{I}(X_i;X_j)\).
    \item Build the complete graph with weights \(\widehat{I}\) and extract the maximum spanning tree (we use Prim’s algorithm).
    \item Choose a root (we pick \(\texttt{height}\), high MI centrality), orient edges away from it, and estimate smoothed parameters for the root marginal and each CPT.
\end{enumerate}

We can now compute full probabilities under the learned model:
\[
\widehat{p}_T(x)=\widehat{p}(h)\;\widehat{p}(w\mid h)\;\widehat{p}(s\mid h)\;\widehat{p}(a\mid h)\;\widehat{p}(g\mid a).
\]

Let's consider a very typical response: 
\begin{align*}
x_{\text{good}}=\{ & \;H_3=\text{tall},\; W_3=\text{heavy},\; S=M,\; \\
& A_2=15\text{--}17,\; G_2=\text{high}\;\}.
\end{align*}
Using the tables above:
\[
\widehat{p}_T(x_{\text{good}})=
\frac{4}{13}\cdot\frac{1}{2}\cdot\frac{3}{5}\cdot\frac{4}{5}\cdot\frac{3}{4}
\;\approx\; 0.055.
\]
Thus \(\log \widehat{p}_T(x_{\text{good}})\approx -2.90\).

But a very atypical case: 
\begin{align*}
x_{\text{bad}}=\{&\;H_1=\text{short},\; W_3=\text{heavy},\; S=F,\; \\
& A_1=12\text{--}14,\; G_2=\text{high}\;\}.
\end{align*}

would  give:
\[
\widehat{p}_T(x_{\text{bad}})=
\frac{3}{13}\cdot\frac{1}{5}\cdot\frac{3}{4}\cdot\frac{3}{4}\cdot\frac{1}{6}
\;\approx\; 0.0043,
\]
so \(\log \widehat{p}_T(x_{\text{bad}})\approx -5.44\), markedly lower than the typical case. 
This concretely shows how violating learned dependencies drives down likelihood.

\section{Statistical Analysis}\label{appendix_more_results}

In Table~\ref{tab:method_correlations}, we report the detailed correlation results computed separately for each method.  
Each cell lists the Pearson correlation coefficient ($r$) and the Spearman rank correlation ($\rho$) between the performance metric (AUC or Lift) and the dataset characteristics (number of samples, variables, total features, and average number of features per variable, AFV).  
The last column in the AUC block additionally reports the correlation between each method’s AUC and its corresponding reconstruction Lift across datasets, indicating whether reconstruction quality aligns with inattentiveness detection ability.  
As shown, most correlations are small and not statistically significant.  
Nevertheless, the Percentile-Loss variant exhibits a significant positive AUC–Lift association, suggesting that models trained with Percentile Loss better align reconstruction fidelity with anomaly sensitivity—consistent with our main trade-off discussion in Section~\ref{subsec:pl_tradeoff}.  
Chow--Liu, as a probabilistic model, has no reconstruction component and thus is excluded from the Lift analysis.

\begin{figure}[t]
    \centering
    \includegraphics[width=0.48\textwidth]{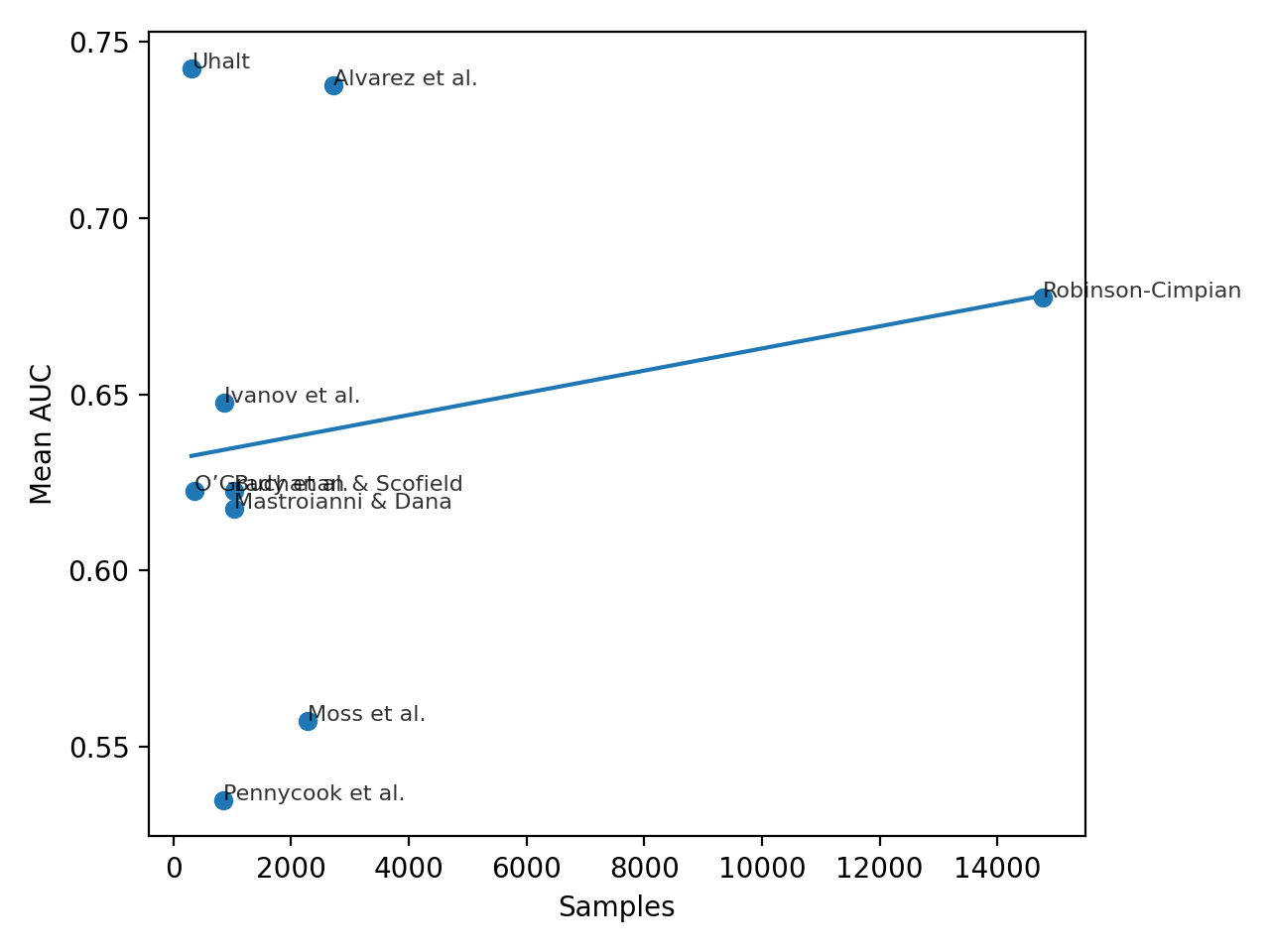}
    \includegraphics[width=0.48\textwidth]{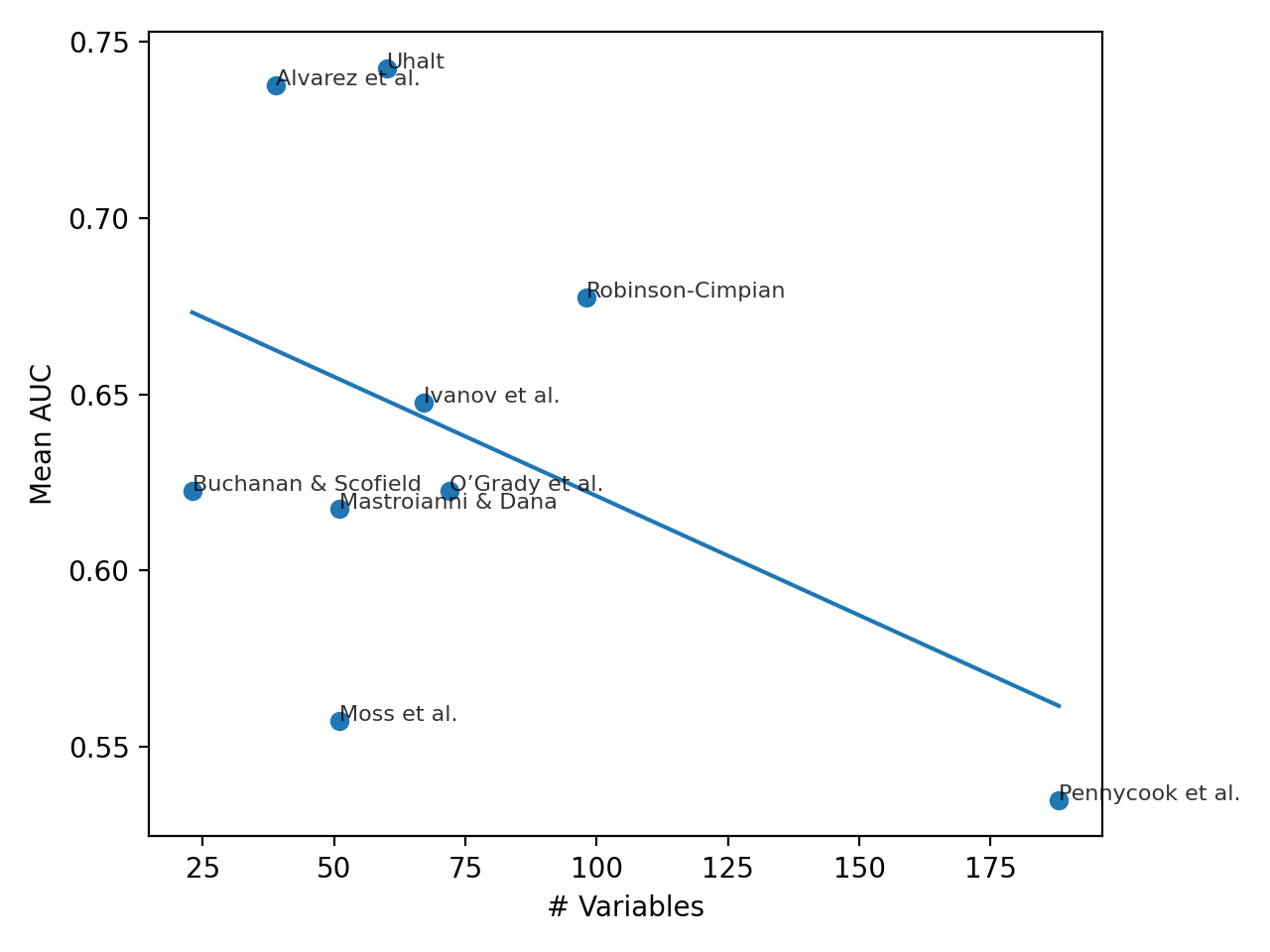}
    \includegraphics[width=0.48\textwidth]{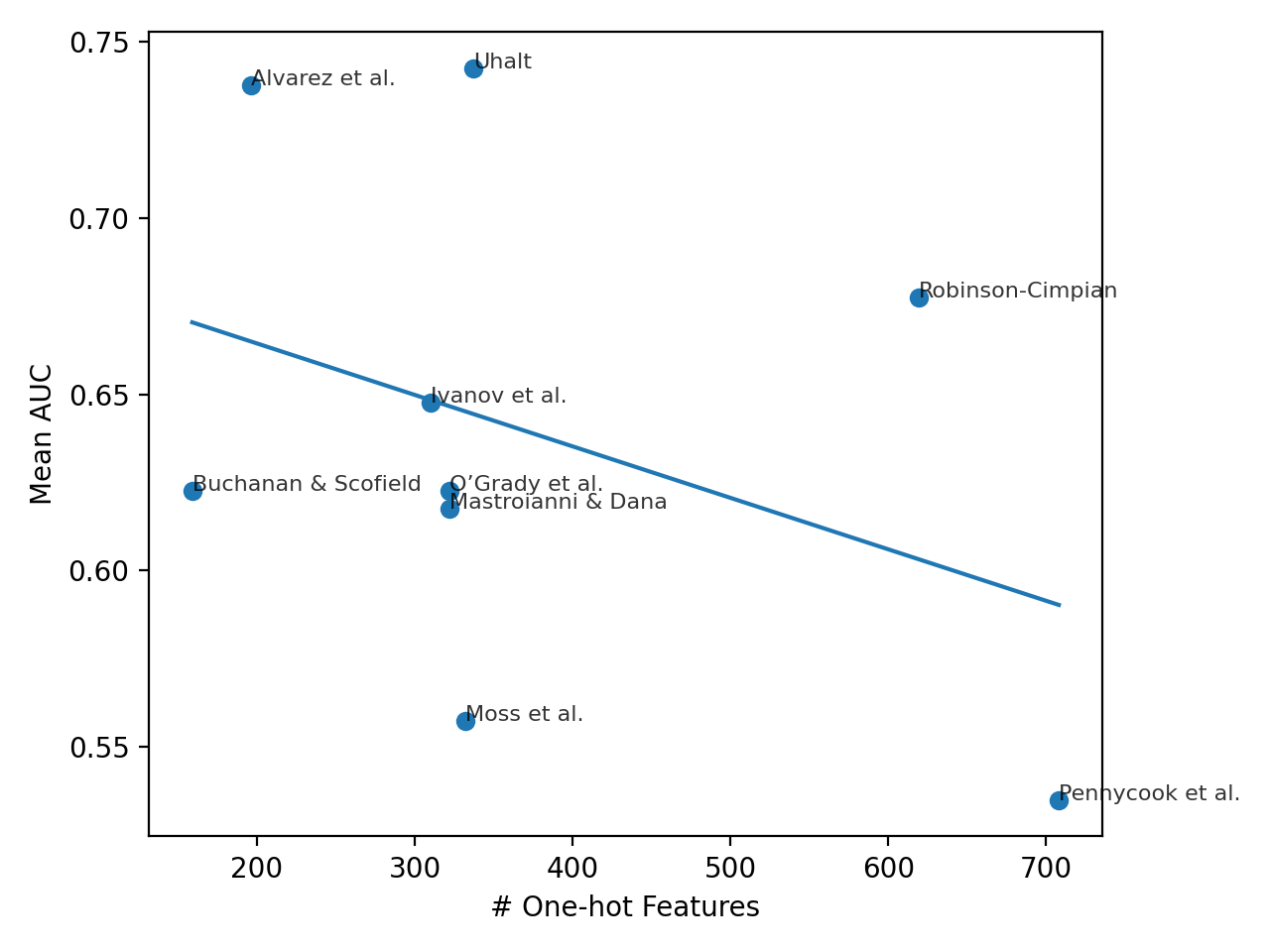}
    \includegraphics[width=0.48\textwidth]{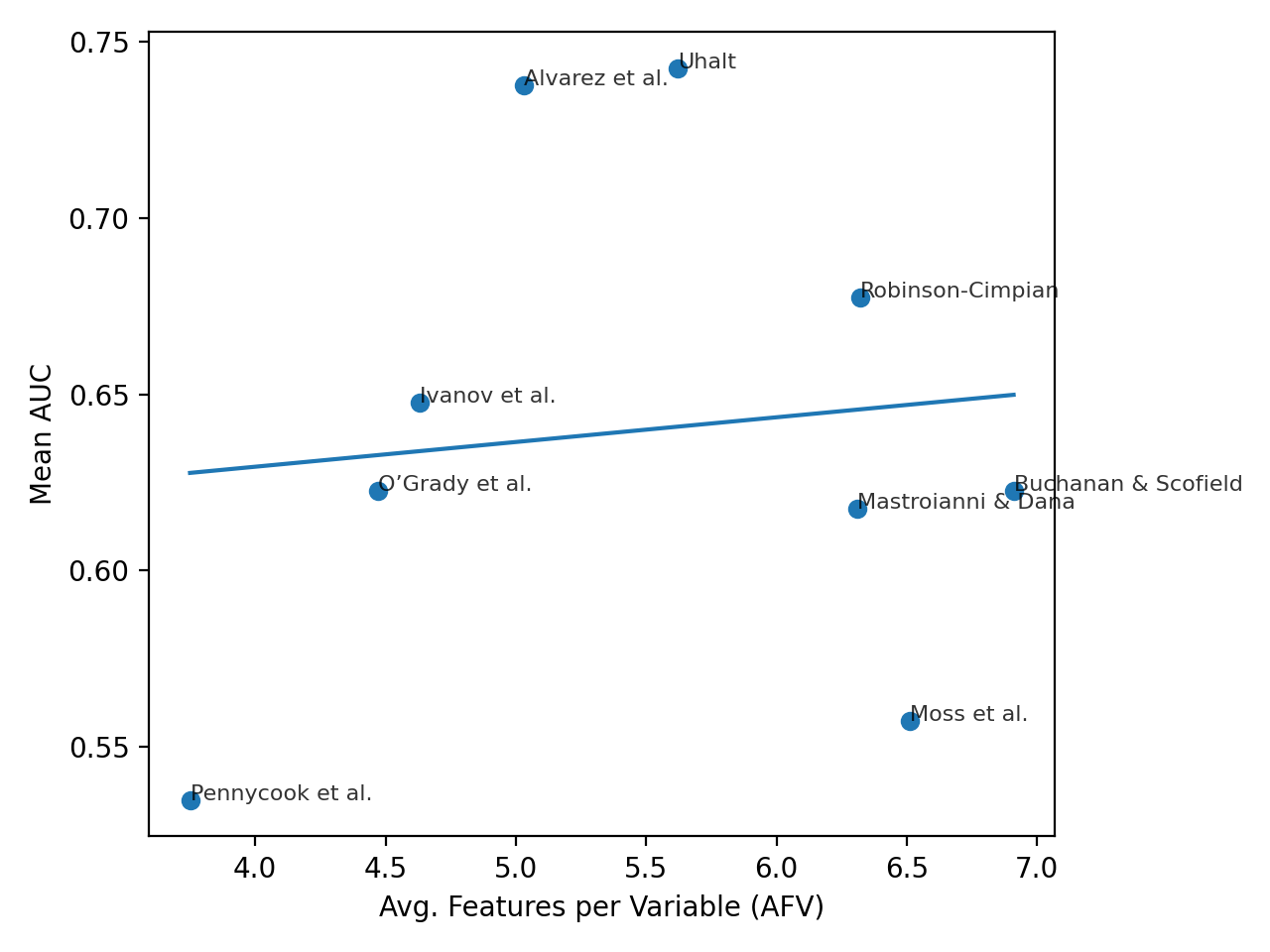}
    \includegraphics[width=0.48\textwidth]{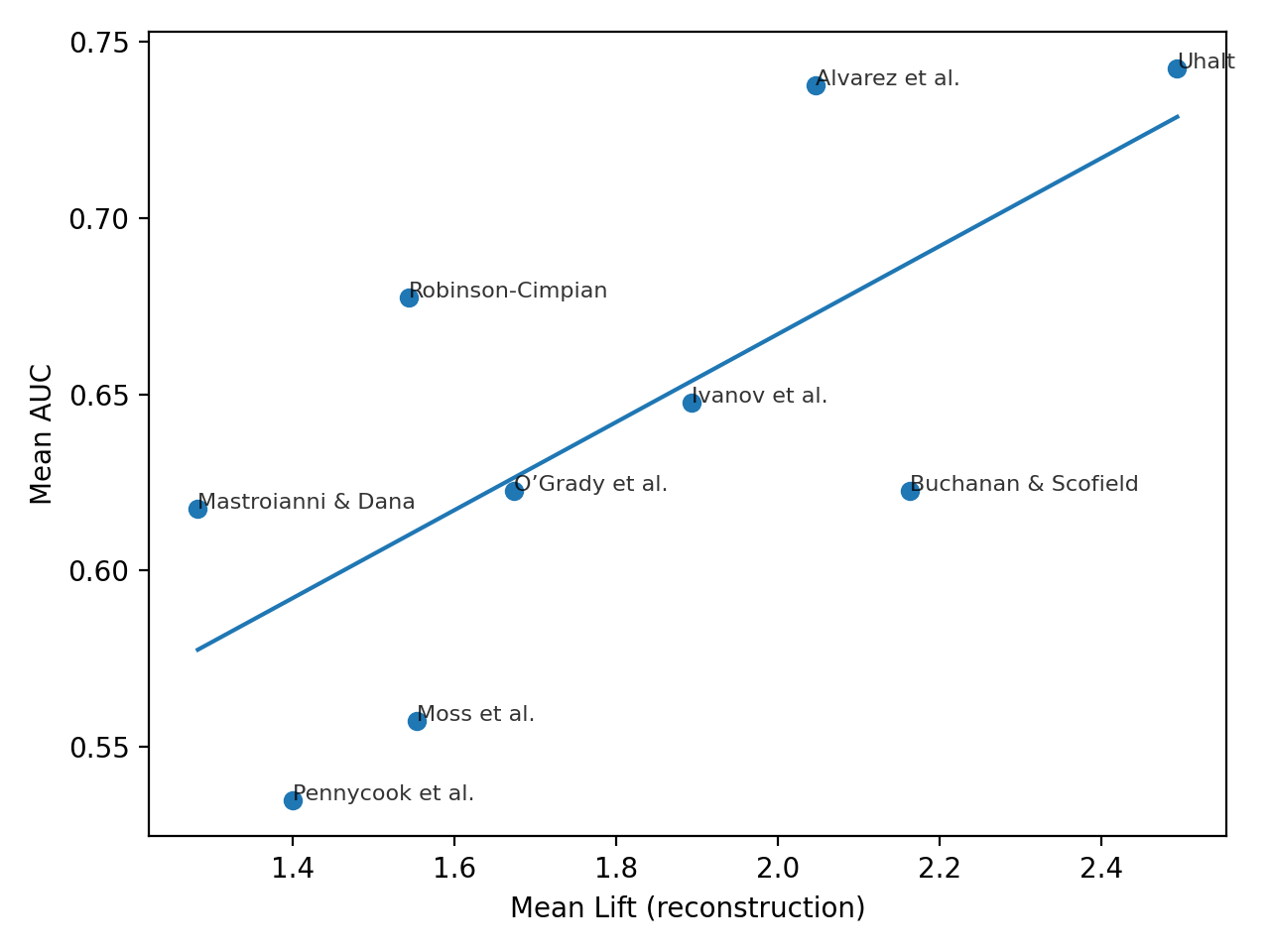}
    \caption{Scatter plots of mean AUC versus dataset characteristics.  
    Each point corresponds to one dataset.  
    The solid line indicates the least-squares regression fit.  
    From top to bottom and left to right: AUC vs.\ Samples, Variables, Features, AFV, and Mean Lift.  
    No strong monotonic trend is evident, confirming that detection performance is largely independent of dataset scale or sparsity, and more strongly driven by latent structural coherence.}
    \label{fig:auc_vs_predictors}
\end{figure}

In Fig.~\ref{fig:auc_vs_predictors}, we illustrate the average AUC performance across different methods for each dataset. Taken together, the correlations and scatter plots reinforce our main conclusion:  
(1) dataset size, dimensionality, and feature granularity have negligible impact on inattentiveness detection performance, and  
(2) datasets where reconstruction models achieve larger relative improvements (higher Lift) tend to yield stronger separability of inattentive respondents.  
These findings support the view that underlying survey structure, not sample count or feature density, is the primary determinant of unsupervised detection success.

\begin{table}[H]
\centering
\resizebox{0.8\textwidth}{!}{
\TABLE
{Method-specific correlations with dataset characteristics \label{tab:method_correlations}}
{\begin{tabular}{lccccccccc}
\toprule
 & \multicolumn{5}{c}{\textbf{AUC}} & \multicolumn{4}{c}{\textbf{Lift}} \\
\cmidrule(lr){2-6}\cmidrule(lr){7-10}
\textbf{Method} & \textbf{Samples} & \textbf{Variables} & \textbf{Features} & \textbf{AFV} & \textbf{Lift (AUC$\leftrightarrow$Lift)} & \textbf{Samples} & \textbf{Variables} & \textbf{Features} & \textbf{AFV} \\
\midrule
Non-Linear AE ($p=100$) 
  & $0.38;\;0.38$
  & $-0.39;\;-0.27$
  & $-0.28;\;-0.23$
  & $0.18;\;0.20$
  & $0.36;\;0.40$
  & $-0.24;\;-0.17$
  & $-0.47;\;-0.39$
  & $-0.54;\;-0.45$
  & $0.10;\;0.13$
\\
Non-Linear AE ($p=80$) 
  & $0.25;\;0.25$
  & $-0.53;\;-0.41$
  & $-0.39;\;-0.25$
  & $0.42;\;0.38$
  & $\mathbf{0.69^{\ast};\;0.72^{\ast}}$
  & $-0.23;\;-0.17$
  & $-0.47;\;-0.39$
  & $-0.55;\;-0.45$
  & $0.10;\;0.13$
\\
Linear AE
  & $-0.34;\;-0.48$
  & $0.04;\;0.10$
  & $-0.11;\;0.22$
  & $-0.45;\;-0.44$
  & $0.27;\;0.18$
  & $-0.26;\;-0.20$
  & $-0.46;\;-0.46$
  & $-0.52;\;-0.49$
  & $0.11;\;0.15$
\\
Chow--Liu Tree
  & $0.13;\;-0.13$
  & $-0.38;\;-0.11$
  & $-0.32;\;-0.22$
  & $-0.06;\;-0.10$
  & \multicolumn{1}{c}{---}
  & \multicolumn{4}{c}{(no reconstruction metric)}
\\
\bottomrule
\end{tabular}}
{Each cell shows \emph{Pearson $r$; Spearman $\rho$} computed across $n{=}9$ datasets (two-sided tests). Asterisks mark statistical significance at $p<.05$. Chow--Liu provides detection AUC but no reconstruction Lift.}}
\end{table}

\section{Trade-Off Plots}
\label{app:tradeoff}

This appendix complements Sec.~\ref{subsec:pl_tradeoff} by visualizing how the Percentile Loss threshold $p$ trades off reconstruction fidelity and inattentiveness detection. Each figure plots the change in a metric relative to the full-batch objective ($p{=}100$), i.e., $\Delta M(p) \equiv M(p) - M(100)$, pooled across datasets/conditions. Consistent with the main text, reconstruction metrics (Accuracy, Lift, ORA) improve monotonically as $p\!\to\!100$, while detection metrics (AUC, NDCG@h, P@10/50/100, R@h) are maximized around $p\in[85,90]$ and taper off near $p{=}95$–$100$.

Figure~\ref{fig:pl_box_acc} shows $\Delta\mathrm{Accuracy}$ vs.\ $p$; medians are negative at $p{=}80/85/90$ and approach $0$ by $p{=}95$. Figure~\ref{fig:pl_box_lift} shows $\Delta\mathrm{Lift}$ with the same monotone pattern toward $p{=}100$. Figure~\ref{fig:pl_box_ora} displays $\Delta\mathrm{ORA}$, again trending upward as $p$ increases. Together, these confirm that PL sacrifices reconstruction at lower $p$ and recovers it when $p$ approaches $100$. See also Sec.~\ref{subsec:pl_tradeoff} for paired tests that quantify these shifts.

Figure~\ref{fig:pl_box_auc} visualizes $\Delta\mathrm{AUC}$. Gains are largest at $p{=}85$ (and robust at $p{=}90$), shrink at $p{=}95$, and vanish or reverse at $p{=}100$. Figures~\ref{fig:pl_box_ndcg}–\ref{fig:pl_box_rh} show similar patterns for NDCG@h, P@10/50/100, and R@h: detection is typically strongest for $p\in[85,90]$, consistent with the sign-tests reported in the main text.

\begin{figure}[p] % 'p' places this on a dedicated page to fit all 9 tightly
  \centering
  \captionsetup{font=small} % Optional: makes caption text slightly smaller for readability
  
  % --- Row 1: Reconstruction Metrics ---
  \begin{subfigure}[t]{0.32\textwidth}
    \centering
    \includegraphics[width=\linewidth]{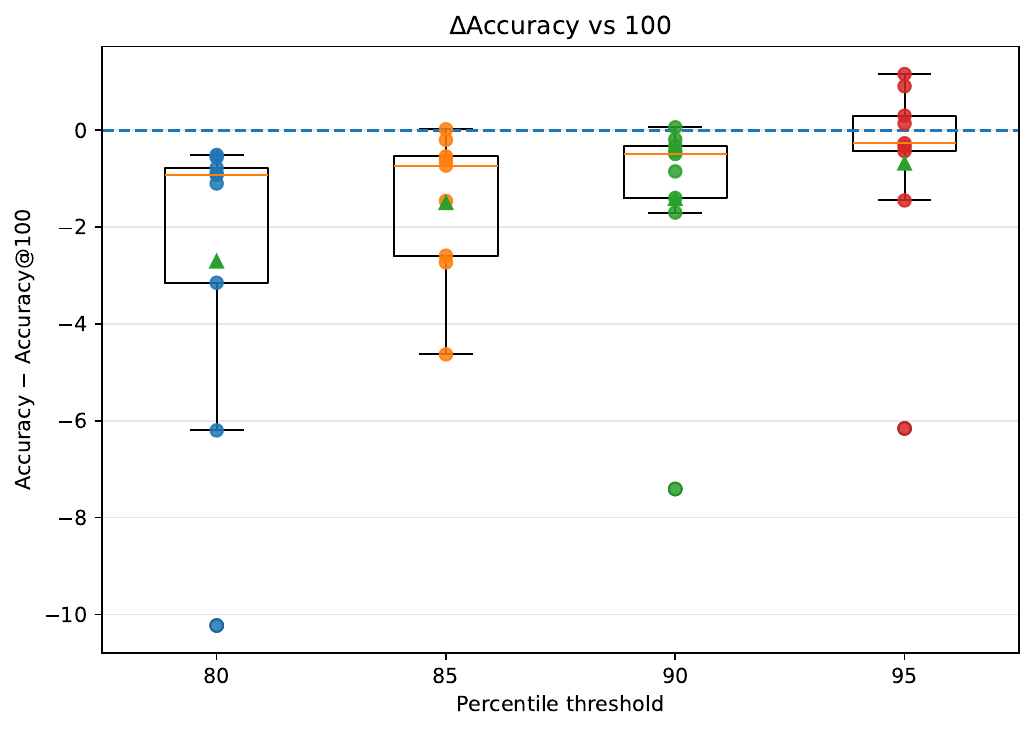}
    \caption{$\Delta$Accuracy}
    \label{fig:pl_box_acc}
  \end{subfigure}
  \hfill
  \begin{subfigure}[t]{0.32\textwidth}
    \centering
    \includegraphics[width=\linewidth]{figures/boxswarm_delta_Lift.pdf}
    \caption{$\Delta$Lift}
    \label{fig:pl_box_lift_app}
  \end{subfigure}
  \hfill
  \begin{subfigure}[t]{0.32\textwidth}
    \centering
    \includegraphics[width=\linewidth]{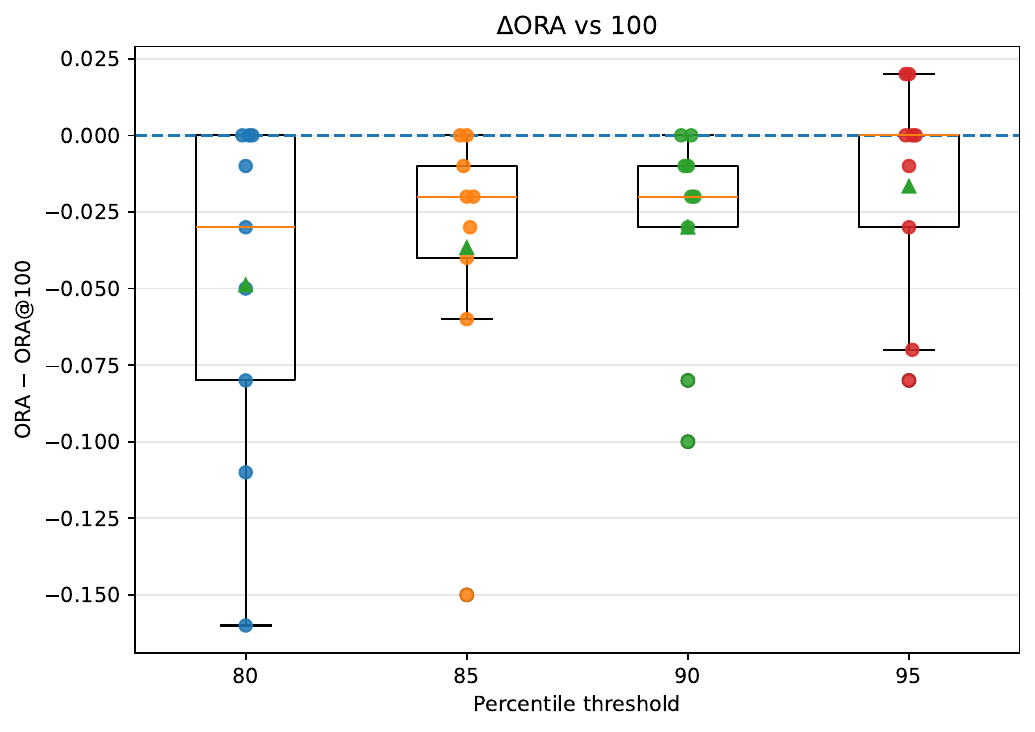}
    \caption{$\Delta$ORA}
    \label{fig:pl_box_ora}
  \end{subfigure}
  
  \vspace{1em} 

  % --- Row 2: Detection Metrics (AUC, NDCG, P@10) ---
  \begin{subfigure}[t]{0.32\textwidth}
    \centering
    \includegraphics[width=\linewidth]{figures/boxswarm_delta_random_AUC.pdf}
    \caption{$\Delta$AUC}
    \label{fig:pl_box_auc_app}
  \end{subfigure}
  \hfill
  \begin{subfigure}[t]{0.32\textwidth}
    \centering
    \includegraphics[width=\linewidth]{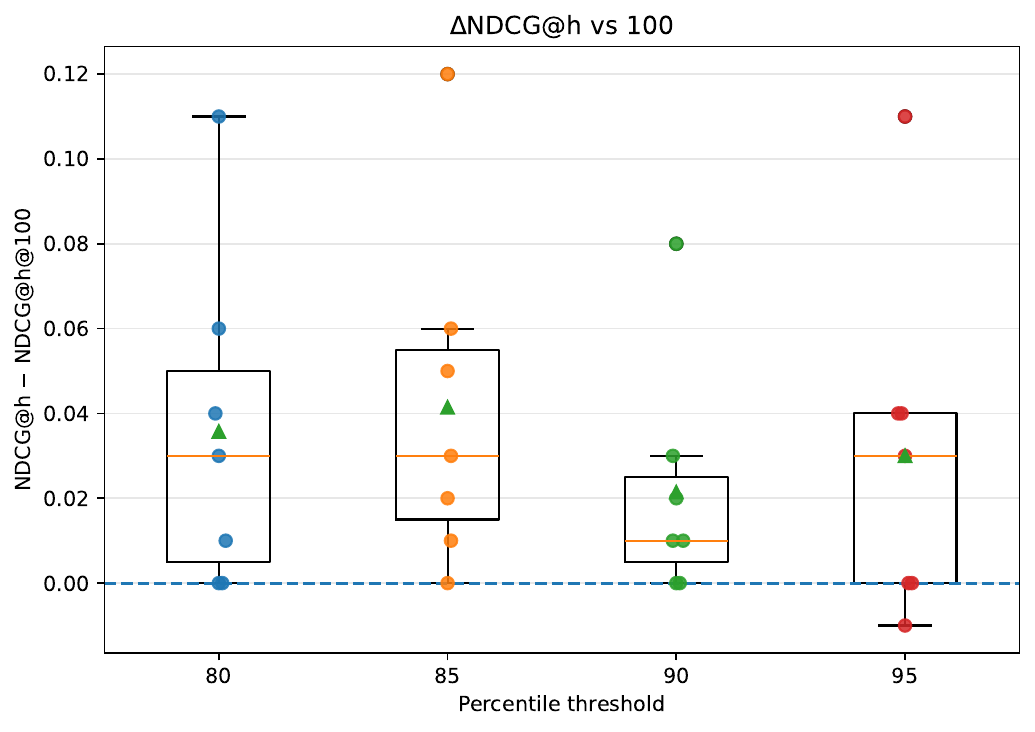}
    \caption{$\Delta$NDCG@h}
    \label{fig:pl_box_ndcg}
  \end{subfigure}
  \hfill
  \begin{subfigure}[t]{0.32\textwidth}
    \centering
    \includegraphics[width=\linewidth]{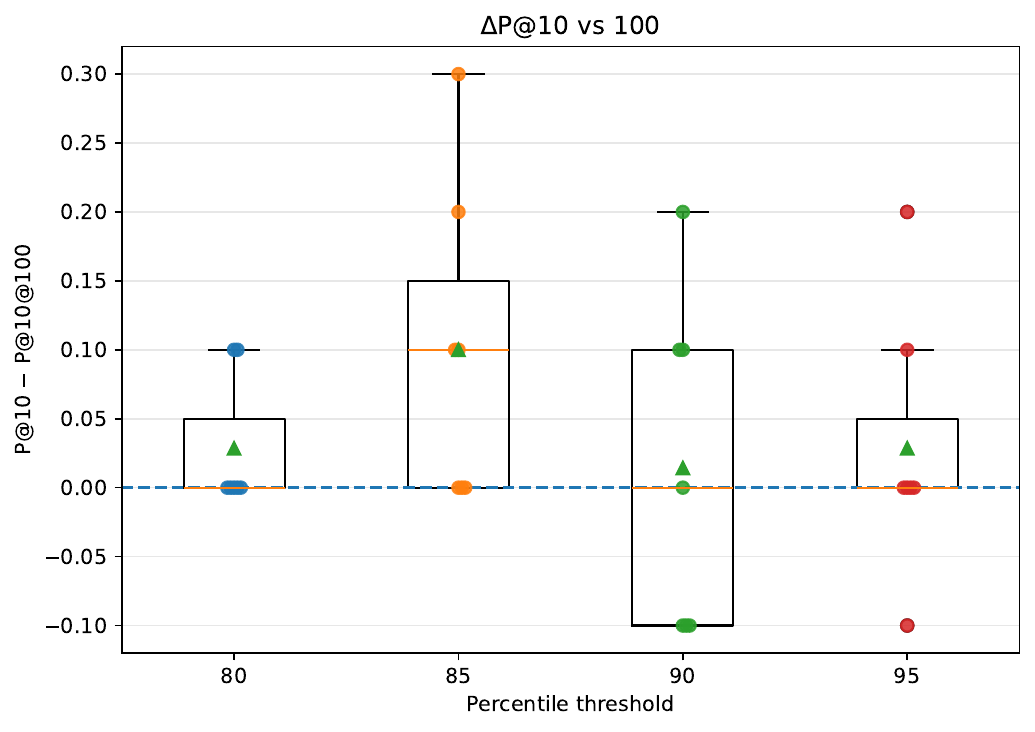}
    \caption{$\Delta$P@10}
    \label{fig:pl_box_p10}
  \end{subfigure}

  \vspace{1em}

  % --- Row 3: Detection Metrics (P@50, P@100, R@h) ---
  \begin{subfigure}[t]{0.32\textwidth}
    \centering
    \includegraphics[width=\linewidth]{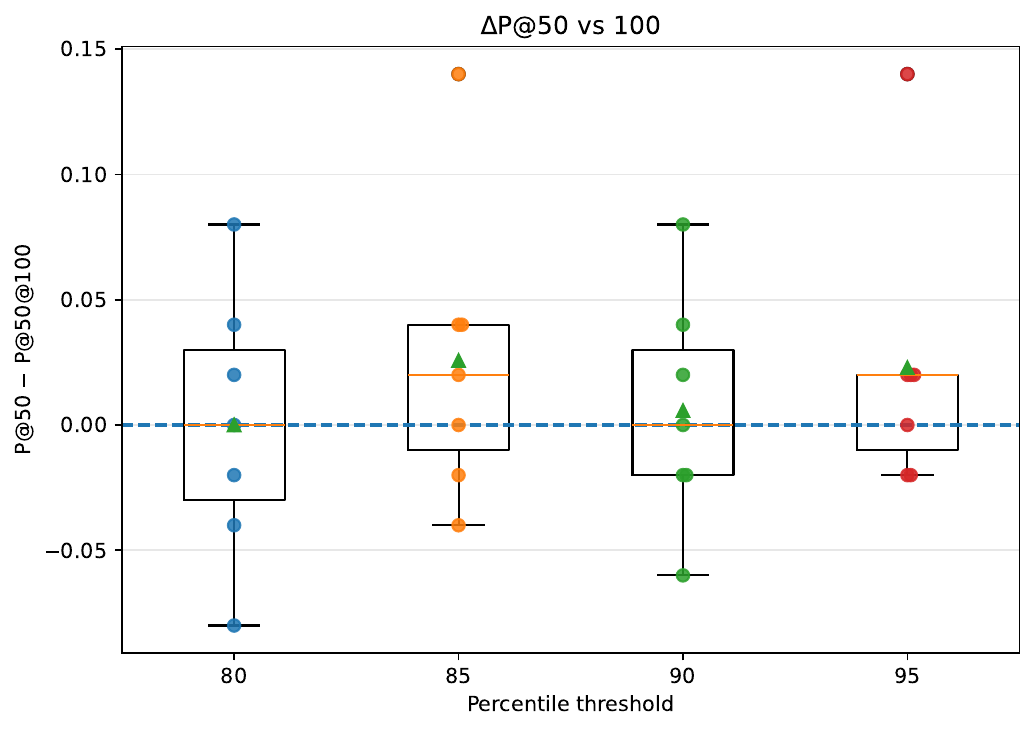}
    \caption{$\Delta$P@50}
    \label{fig:pl_box_p50}
  \end{subfigure}
  \hfill
  \begin{subfigure}[t]{0.32\textwidth}
    \centering
    \includegraphics[width=\linewidth]{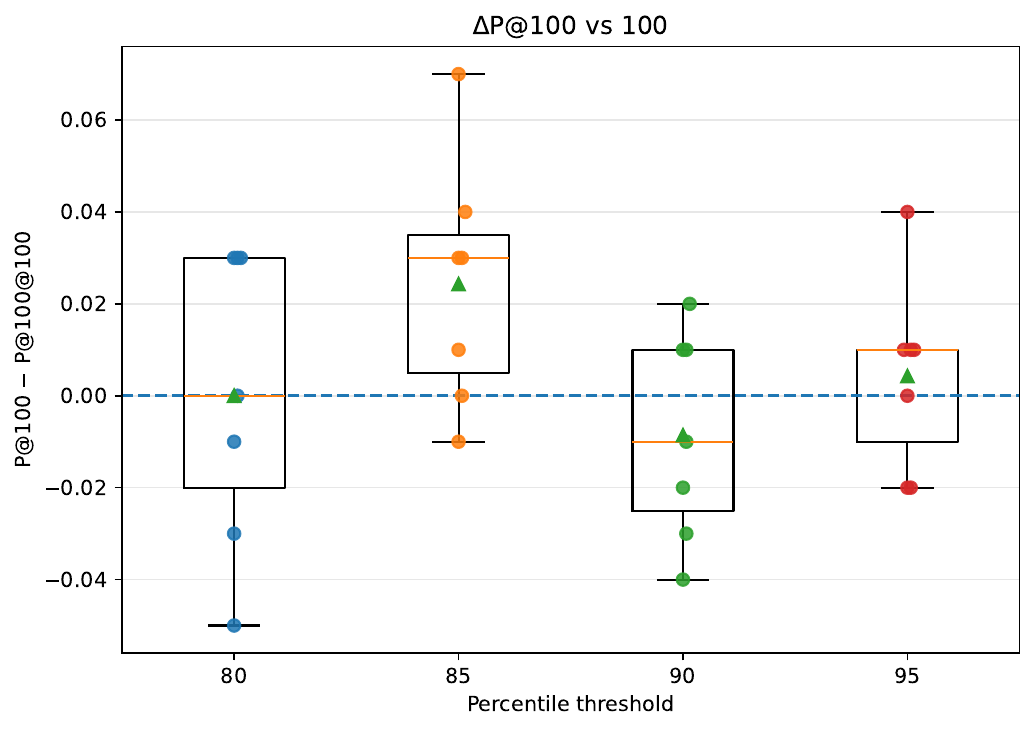}
    \caption{$\Delta$P@100}
    \label{fig:pl_box_p100}
  \end{subfigure}
  \hfill
  \begin{subfigure}[t]{0.32\textwidth}
    \centering
    \includegraphics[width=\linewidth]{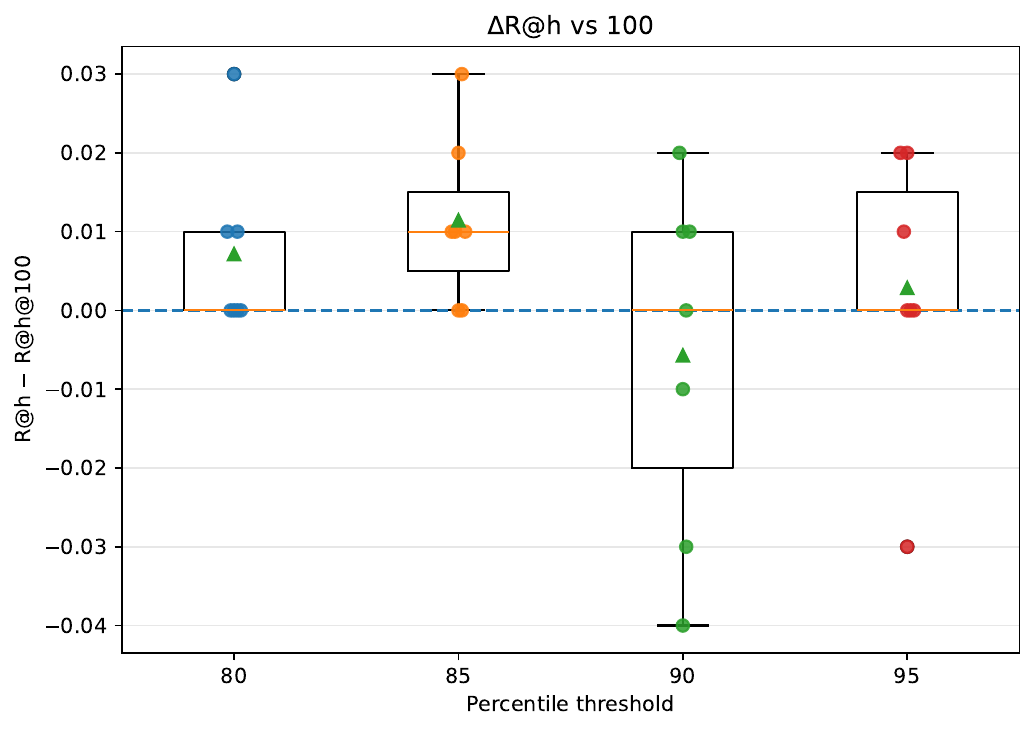}
    \caption{$\Delta$R@h}
    \label{fig:pl_box_rh}
  \end{subfigure}

  \caption{Reconstruction and Detection trade-offs relative to $p{=}100$. 
  (a) Medians are negative at $p{=}80/85/90$ and approach $0$ at $p{=}95$. 
  (b) Reconstruction improves as $p$ increases. 
  (c) Higher $p$ yields better ranking calibration for reconstruction. 
  (d) Broad optimum around $p{\approx}85$--$90$. 
  (e) Ranking gains mirror AUC, peaking at $p{=}85$--$90$. 
  (f) Precision at short lists benefits most from $p{=}85$--$90$. 
  (g) Improvements persist but are smaller than at P@10. 
  (h) Gains are modest, consistent with the broader cut. 
  (i) Recall improvements peak around $p{=}85$--$90$.}
  \label{fig:all_tradeoffs}
\end{figure}

\end{APPENDICES}